\documentclass{iopart}
\usepackage{graphicx,bm,amssymb,iopams}

\eqnobysec
\newcommand{\pdag}{{\phantom{\dagger}}}
\newcommand{\pprime}{{\phantom{\prime}}}
\newcommand{\past}{{\phantom{\ast}}}

\newcommand{\ket}[1]{\left|{#1}\right\rangle}

\newcommand{\PRB}{{\it Phys. Rev. B~}} 

\begin{document}

\title{Kondo effect in quantum dots}

\author{Michael Pustilnik\dag\ ~and~ Leonid Glazman\ddag}

\address{\dag\ School of Physics, Georgia Institute of Technology,
Atlanta, GA 30332, USA}
\address{\ddag\ William I. Fine Theoretical Physics Institute, University of Minnesota, 
Minneapolis, MN 55455, USA}


\begin{abstract}
We review mechanisms of low-temperature electronic transport 
through a quantum dot weakly coupled to two conducting leads. 
Transport in this case is dominated by electron-electron 
interaction. At temperatures moderately lower than the charging 
energy of the dot, the linear conductance is suppressed by the 
Coulomb blockade. Upon further lowering of the temperature, 
however, the conductance may start to increase again due to the 
Kondo effect. We concentrate on lateral quantum dot systems 
and discuss the conductance in a broad temperature range, which 
includes the Kondo regime. 
\end{abstract}


\section{Introduction}

In quantum dot devices~\cite{blockade} a small droplet of 
electron liquid is confined in a finite region of space. The 
droplet can be attached by tunneling junctions to massive 
electrodes to allow electronic transport across the system. 
The conductance of such a device is determined by the number 
of electrons on the dot $N$, which in turn is controlled by 
varying the potential on the gate - an auxiliary electrode 
capacitively coupled to the dot~\cite{blockade}. At sufficiently 
low temperatures the number of electrons $N$ is an integer at 
almost any gate voltage $V_g$. Exceptions are narrow intervals 
of $V_g$ in which an addition of a single electron to the dot does 
not change significantly the electrostatic energy of the system. Such a 
degeneracy between different charge states of the dot allows for 
an activationless electron transfer through it, whereas for all other 
values of $V_g$ the activation energy for the conductance $G$ 
across the dot is finite. The resulting oscillatory dependence 
$G(V_g)$ is the hallmark of the Coulomb blockade 
phenomenon~\cite{blockade}. The contrast between the low- 
and high-conductance regions (Coulomb blockade valleys and 
peaks, respectively) gets sharper at lower temperatures. This 
pattern of $G(V_g,T)$ dependence is observed down to the 
lowest attainable temperatures in experiments on tunneling through 
small metallic islands~\cite{Devoret}. However, small quantum dots 
formed in GaAs heterostructures exhibit drastically different 
behavior~\cite{kondo_exp}: in some Coulomb blockade valleys 
the dependence $G(T)$ is not monotonic and has a minimum 
at a finite temperature. This minimum is similar in 
origin~\cite{kondo_popular} to the well-known non-monotonic 
temperature dependence of the resistivity of a metal containing 
magnetic impurities~\cite{Kondo} -- the \textit{Kondo effect}.

In this paper we review the theory of the Kondo effect in quantum 
dots, concentrating on the so-called \textit{lateral quantum dot 
systems}~\cite{blockade,kondo_exp}, formed by gate depletion 
of a two-dimensional electron gas at the interface between 
two semiconductors. These devices offer the highest degree 
of tunability, yet allow for relatively simple theoretical treatment. 
At the same time, many of the results presented below are directly 
applicable to other systems as well, including  vertical quantum 
dots~\cite{vertical,Sasaki,induced_review}, Coulomb-blockaded 
carbon nanotubes~\cite{induced_review,nanotube}, single-molecule 
transistors~\cite{Park}, and stand-alone magnetic atoms on 
metallic surfaces~\cite{mirage}. 

\section{Model of a lateral quantum dot system} 
\label{CIM}

The Hamiltonian of interacting electrons confined to a quantum dot 
has the following general form,
\begin{equation}
H_{\rm dot} = 
\sum_s \sum_{ij}
h^\pdag_{ij}d^\dagger_{is}d^\pdag_{js}
+
\frac{1}{2}\sum_{s s'}\sum_{ijkl} h^\pdag_{ijkl}
d^\dagger_{i s} d^\dagger_{j s'} 
d^\pdag_{k s'}d^\pdag_{l s}.
\label{2.1}
\end{equation}
Here an operator $d^\dagger_{is}$ creates an electron with spin 
$s$ in the orbital state $\phi_i(\bi r)$; $h_{ij}^\past=h^*_{ji}$ 
is an Hermitian matrix describing the single-particle part of the 
Hamiltonian. The matrix elements $h_{ijkl}$ depend on the 
potential $U(\bi r-\bi r')$ of electron-electron interaction,
\begin{equation}
h_{ijkl}
=\int d\bi r\, d\bi r' 
\phi_i^*(\bi r) \phi_j^*(\bi r')
U(\bi r-\bi r')
\phi_k^\past (\bi r') \phi_l^\past (\bi r).
\label{2.2}
\end{equation}

The Hamiltonian~\eref{2.1} can be simplified further provided 
that the quasiparticle spectrum is not degenerate near the Fermi 
level, that the Fermi-liquid theory is applicable to the description 
of the dot, and that the dot is in the metallic conduction regime. 
The first of these conditions is satisfied if the dot has no spatial 
symmetries, which implies also that motion of quasiparticles within 
the dot is chaotic. 

The second condition is met if the electron-electron interaction 
within the dot is not too strong, i.e. the gas parameter $r_s$ 
is small, 
\begin{equation}
r_s = (k_F a_0)^{-1} \lesssim 1,
\quad
a_0 = \kappa \hbar^2/e^2 m^\ast 
\label{2.3}
\end{equation}
Here $k_F$ is the Fermi wave vector, $a_0$ is the effective 
Bohr radius, $\kappa$ is the dielectric constant of the material, 
and $m^\ast$ is the quasiparticle effective mass.

The third condition requires the ratio of the Thouless energy 
$E_T$ to the mean single-particle level spacing $\delta E$ to 
be large~\cite{RMT1}, 
\begin{equation}
g = E_T/\delta E \gg 1.
\label{2.4}
\end{equation}
For a ballistic two-dimensional dot of linear size $L$ the Thouless 
energy $E_T$ is of the order of $\hbar v_F/L$, whereas the level 
spacing can be estimated as 
\begin{equation}
\delta E \sim \hbar v_F k_F/N \sim \hbar^2/m^\ast L^2 .
\label{2.5}
\end{equation}
Here $v_F$ is the Fermi velocity and $N\sim (k_F L)^2$ is the 
number of electrons in the dot. Therefore,
\[
g \sim k_F L \sim \sqrt N\,,
\]
so that having a large number of electrons $N\gg 1$ in the dot 
guarantees that the condition~\eref{2.4} is satisfied.  

Under the conditions~\eref{2.3},~\eref{2.4} the \textit{Random 
Matrix Theory} (see~\cite{RMT_reviews,Mehta} for a review), 
is a good starting point for description of non-interacting 
quasiparticles within the energy strip of width $E_T$ about 
the Fermi level~\cite{RMT1}. The matrix elements $h_{ij}$ in 
Eq.~\eref{2.1} belong to a Gaussian ensemble~\cite{Mehta}. Since 
the matrix elements do not depend on spin, each eigenvalue 
$\epsilon_n$ of the matrix $h_{ij}$ represents a spin-degenerate 
energy level. The spacings $\epsilon_{n+1}-\epsilon_n$ between 
consecutive levels obey the Wigner-Dyson statistics~\cite{Mehta}; 
the mean level spacing 
$\overline{\epsilon_{n+1}-\epsilon_n} = \delta E$.

We discuss now the second term in the Hamiltonian~\eref{2.1}, 
which describes electron-electron interaction. It turns 
out~\cite{RMT,KAA,ABG} that the vast majority of the matrix 
elements $h_{ijkl}$ are small. Indeed, in the lowest order in 
$1/g\ll 1$, the wave functions $\phi_i (\bi r)$ are Gaussian 
random variables statistically independent of each other and 
of the corresponding energy levels~\cite{wave_functions}:
\begin{equation}
\overline {\phi_i^*(\bi r)\phi_j^\past(\bi r')} 
= \frac{1}{L^2} \delta_{ij} F(|\bi r - \bi r'|),
\quad
F(r) \sim \langle\exp(\rmi \bi k\cdot\bi r)\rangle_{\rm FS} 
\label{2.6}
\end{equation}
Here $\langle\ldots\rangle_{\rm FS}$ stands for the averaging 
over the Fermi surface $|\bi k| = k_F$. In two dimensions, the 
function $F(r)$ decreases with $r$ as $F\propto (k_F r)^{-1/2}$ 
at $k_F r\gg 1$, and saturates to $F\sim 1$ at $k_F r\ll 1$. 
After averaging with the help of Eq.~\eref{2.6}, the matrix
elements~\eref{2.2} take the form\footnote[1]{For simplicity 
we assumed here that 
$\overline {\phi_i(\bi r)\phi_j(\bi r')} \equiv 0$, 
which corresponds to broken time-reversal symmetry. 
See~\cite{ABG} for discussion of the general case.}
\[
\overline{h_{ijkl}} 
= 2E_C\delta_{il}\delta_{jk} + E_S\delta_{ik}\delta_{jl} .
\]
We substitute this expression into the Hamiltonian~\eref{2.1}, and 
rearrange the sum over the spin indices with the help of the identity 
\begin{equation}
2\,\delta_{s_1^\pprime s_2^\pprime} \delta_{s_1^\prime s_2^\prime} 
= \delta_{s_1^\pprime s_1^\prime} \delta_{s_2^\pprime s_2^\prime} 
+ \bsigma_{s_1^\pprime s_1^\prime} \cdot
\bsigma_{s_2^\pprime s_2^\prime} ,
\label{0}
\end{equation}
where $\bsigma = (\sigma^x,\sigma^y,\sigma^z)$ are the Pauli 
matrices. This results in a remarkably simple form~\cite{KAA,ABG}
\begin{equation}
H_{\rm int} = E_C\hat{N}^2 - E_S \hat{\bf S}^2
\label{2.7}
\end{equation}
of the interaction part of the Hamiltonian of the dot. Here 
\begin{equation}
 \hat{N} =\sum_{ns} d^\dagger_{ns}d^\pdag_{ns},
\quad
\hat{\bf S} = \sum_{nss'}
d^\dagger_{ns}\frac{{\bm\sigma}_{ss'}}{2}d^\pdag_{ns'}
\label{2.8}
\end{equation}
are the operators of the total number of electrons in the dot and 
of the dot's spin, respectively.    

\begin{figure}[h]
\centerline{\includegraphics[width=0.45\textwidth]{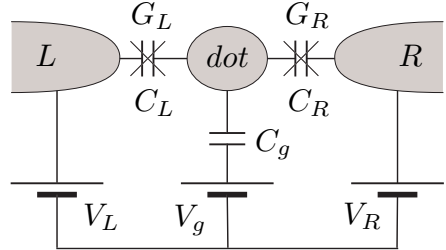}}
\vspace{2.5 mm}
\caption{Equivalent circuit for a quantum dot connected to two 
leads by tunneling junctions and coupled via a capacitor (with 
capacitance $C_g$) to the gate. The total capacitance of the 
dot $C=C_L+C_R+C_g$.}
\label{circuit}
\end{figure}

The first term in Eq.~\eref{2.7} represents the electrostatic energy. 
In the conventional equivalent circuit picture, see Fig.~\ref{circuit}, 
the charging energy $E_C$ is related to the total capacitance $C$ 
of the dot, $E_C=e^2/2C$. For a mesoscopic ($k_F L\gg 1$) 
conductor, the charging energy is large compared to the mean level 
spacing $\delta E$. Indeed, using the estimates $C\sim\kappa L$ 
and~\eref{2.5}, we find
\begin{equation}
E_C/\delta E \sim L/a_0 
\sim r_s \sqrt N .
\label{2.9}
\end{equation}
Except an exotic case of an extremely weak interaction, this ratio is 
large for $N\gg 1$; for the smallest quantum dots formed in GaAs 
heterostructures, $E_C /\delta E\sim 10$~\cite{kondo_exp}. 

The second term in Eq.~\eref{2.7} describes the intra-dot exchange 
interaction, with the exchange energy $E_S$ given by
\begin{equation}
E_S = \int d\bi r\, d\bi r' U(\bi r-\bi r') F^2(|\bi r - \bi r'|) 
\label{2.10}
\end{equation}
In the case of a long-range interaction the potential $U$ here 
should properly account for the screening~\cite{ABG}. For 
$r_s\ll 1$ the exchange energy can be estimated with logarithmic 
accuracy by substituting $U(r) = (e^2/\kappa r)\theta(a_0 - r)$ 
into Eq.~\eref{2.10} (here we took into account that the screening 
length in two dimensions coincides with the Bohr radius $a_0$), 
which yields
\begin{equation}
E_S \sim r_s\ln\left(1/r_s\right)\delta E \ll \delta E. 
\label{2.11}
\end{equation}
The estimate~\eref{2.11} is valid only for $r_s\ll 1$. However, 
the ratio $E_S/\delta E$ remains small for experimentally
relevant\footnote{For GaAs ($m^*\approx 0.07 m_e$, 
$\kappa\approx 13$) the effective Bohr radius $a_0\approx 10~nm$, 
whereas a typical density of the two-dimensional electron gas,
$n\sim10^{11}~cm^{-2}$~\cite{kondo_exp}, corresponds to 
$k_F = \sqrt{2\pi n}\sim 10^6~cm^{-1}$. This gives 
$k_F a_0 \sim 1$.}
value $r_s\sim 1$ as long as the Stoner criterion for the 
absence of itinerant magnetism~\cite{Ziman} is satisfied. 
This guarantees the absence of a macroscopic (proportional 
to $N$) magnetization of a dot in the ground state~\cite{KAA}.

Obviously, the interaction part of the Hamiltonian, Eq.~\eref{2.7}, 
is invariant with respect to a change of the basis of single-particle 
states $\phi_i(\bi r)$. Picking up the basis in which the first term 
in \eref{2.1} is diagonal, we arrive at the \textit{universal 
Hamiltonian}~\cite{KAA,ABG},
\begin{equation}
H_{\rm dot}= 
\sum_{ns}\epsilon^\pdag_n d_{ns}^\dagger d^\pdag_{ns}
+ E_C \bigl({\hat N}- N_0\bigr)^2 -E_S \hat{\bf S}^2.
\label{2.12}
\end{equation}
We included in Eq.~\eref{2.12} the effect of the capacitive coupling 
to the gate electrode: the dimensionless parameter $N_0$ is 
proportional to the gate voltage, $N_0=C_gV_g/e$, where $C_g$ 
is the capacitance between the dot and the gate, see Fig.~\ref{circuit}.  
The relative magnitude of various terms not included in~\eref{2.12}, 
as well as that of mesoscopic fluctuations of the coupling constants 
$E_C$ and $E_S$, is of the order of $1/g\sim N^{-1/2}\ll 1$. 

As discussed above, in this limit the energy scales involved 
in~\eref{2.12} form a well-defined hierarchy 
\begin{equation}
E_S\ll\delta E\ll E_C .
\label{2.13}
\end{equation}
If all the single-particle energy levels $\epsilon_n$ were equidistant, 
then the spin $S$ of an even-$N$ state would be zero, while an 
odd-$N$ state would have $S=1/2$. However, the level spacings 
are random. If the spacing between the highest occupied level and 
the lowest unoccupied one is accidentally small, than the gain in the 
exchange energy, associated with the formation of a higher-spin state, 
may be sufficient to overcome the loss of the kinetic energy (cf. the 
Hund's rule in quantum mechanics). For $E_S\ll \delta E$ such 
deviations from the simple even-odd periodicity are 
rare~\cite{KAA,spin,spin_exp}. This is why the last term 
in~\eref{2.12} is often neglected. \Eref{2.12} then reduces to the 
Hamiltonian of the \textit{Constant Interaction Model}, widely used 
in the analysis of experimental data~\cite{blockade}.

\begin{figure}[h]
\centerline{\includegraphics[width=0.3\textwidth]{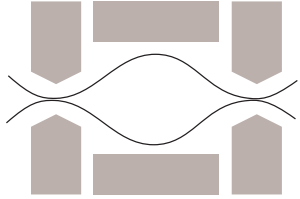}}
\vspace{2.5 mm}
\caption{The confining potential forming a lateral quantum dot varies 
smoothly on the scale of the de Broglie wavelength at the Fermi energy. 
Hence, the dot-lead junctions act essentially as electronic waveguides 
with a well-defined number of propagating modes.}
\label{lateral_dot}
\end{figure}

Electron transport through the dot occurs via two dot-lead junctions.
In a typical geometry, the potential forming a lateral quantum dot varies 
smoothly on the scale of the Fermi wavelength, see Fig.~\ref{lateral_dot}.
Hence, the point contacts connecting the quantum dot to the leads act 
essentially as electronic waveguides. Potentials on the gates control 
the waveguide width, and, therefore, the number of electronic modes 
the waveguide support: by making the waveguide narrower one pinches 
the propagating modes off one-by-one. Each such mode contributes 
$2e^2/h$ to the conductance of a contact. The Coulomb blockade 
develops when the conductances of the contacts are small compared 
to $2e^2/h$, i.e. when the very last propagating mode approaches its 
pinch-off~\cite{KM,KF}.  Accordingly, in the Coulomb blockade 
regime each dot-lead junction in a lateral quantum dot system supports 
only a single electronic mode~\cite{real}.

As discussed below, for $E_C\gg\delta E$ the characteristic energy 
scale relevant to the Kondo effect, the Kondo temperature $T_K$, 
is small compared to the mean level spacing: $T_K\ll\delta E$. This 
separation of the energy scales allows us to simplify the problem 
even further by assuming that the conductances of the dot-lead 
junctions are small. This assumption will not affect the properties 
of the system in the Kondo regime. At the same time, it justifies the 
use of the tunneling Hamiltonian for description of the coupling 
between the dot and the leads. The microscopic Hamiltonian of the 
system can then be written as a sum of three distinct terms,
\begin{equation}
H = H_{\rm leads} + H_{\rm dot} + H_{\rm tunneling},
\label{2.14}
\end{equation}
which describe free electrons in the leads, an isolated quantum dot, 
and tunneling between the dot and the leads, respectively. The 
second term in~\eref{2.14}, the Hamiltonian of the dot 
$H_{\rm dot}$, is given by Eq.~\eref{2.12}. We treat the leads as 
reservoirs of free electrons with continuous spectra $\xi_{k}$, 
characterized by constant density of states $\nu$, same for both 
leads. Moreover, since the typical energies $\omega\lesssim E_C$ 
of electrons participating in transport through a quantum dot in 
the Coulomb blockade regime are small compared to the Fermi 
energy of the electron gas in the leads, the spectra $\xi_{k}$ can 
be linearized near the Fermi level, $\xi_k = v_F k$; here $k$ is 
measured from $k_F$. With only one electronic mode per junction 
taken into account, the first and the third terms in Eq.~\eref{2.14} have 
the form
\begin{eqnarray}
H_{\rm leads} =  \sum_{\alpha ks}\xi^\pdag_k 
c^\dagger_{\alpha ks} c^\pdag_{\alpha ks} ,
\quad
\xi_k = - \xi_{-k},
\label{2.15} \\
H_{\rm tunneling} 
= \sum_{\alpha k n s}t^\pdag_{\alpha n}
c^\dagger_{\alpha k s} d^\pdag_{ns} + {\rm H.c.}
\label{2.16}
\end{eqnarray}
Here $t_{\alpha n}$ are tunneling matrix elements (tunneling 
amplitudes) ``connecting'' the state $n$ in the dot with the state 
$k$ in the lead $\alpha$ ($\alpha =R,L$ for the right/left lead). 
The randomness of states $n$ translates into the randomness 
of the tunneling amplitudes. Indeed, the amplitudes depend on 
the values of the electron wave functions at the points $\bi r_\alpha$ 
of the contacts, $t_{\alpha n}\propto \phi_n(\bi r_\alpha)$. For 
$k_F L\gg 1$ the wave functions are Gaussian random variables. 
\Eref{2.6} then results in
\begin{equation}
\overline{t^\ast_{\alpha n} t^\past_{\alpha'n'}} 
= \overline{\left|t^2_{\alpha n}\right|} 
\delta^\past_{\alpha\alpha'} \delta^\past_{nn'} .
\label{2.17}
\end{equation}
Average values of the tunneling probabilities can be expressed 
via the conductances of the dot-lead junctions $G_\alpha$,
\begin{equation}
\frac{h}{2e^2}\,G_\alpha 
= \frac{\Gamma_\alpha}{\delta E} 
\sim \frac{\nu\overline{\left|t^2_{\alpha n}\right|}}{\delta E} \,.
\label{2.18}
\end{equation}
Here $\Gamma_\alpha$ is the average rate for an electron to 
escape from a discrete level $n$ in the dot into lead $\alpha$. 

\section{Rate equations and conductance across the dot}

At high temperatures, $T\gg E_C $, charging energy is 
negligible compared to the thermal energy of electrons. 
Therefore the conductance of the device in this regime 
$G_\infty$ is not affected by charging and, independently 
of the gate voltage,
\begin{equation}
\frac{1}{\,G_\infty}=\frac{1}{G_L}+\frac{1}{G_R}.
\label{3.1}
\end{equation}
Dependence on $N_0$ develops at lower temperatures,
\begin{equation}
\delta E\ll T\ll E_C. 
\label{3.2}
\end{equation}
The conductance is not suppressed only within narrow regions 
-- \textit{Coulomb blockade peaks}, i.e. when the gate voltage 
is tuned sufficiently close to one of the points of charge 
degeneracy,
\begin{equation}
|N_0^\past - N_0^*|\lesssim T/E_C ;
\label{3.3}
\end{equation}
here $N_0^*$ is a half-integer number. 

We will demonstrate this now using the method of rate 
equations~\cite{rate}. In addition to constraint~\eref{3.2}, 
we will assume that the inelastic electron relaxation rate within 
the dot is large compared to the escape rates $\Gamma_\alpha$. 
In other words, transitions between discrete levels in the dot 
occur before the electron escapes to the 
leads\footnote{Note that a finite inelastic relaxation rate requires 
inclusion of mechanisms beyond the model~\eref{2.12}, e.g., 
electron-phonon collisions.}. Under this assumption the 
tunnelings across the two junctions can be treated independently 
of each other. Condition~\eref{3.3}, on the other hand, 
allows us to take into account only two charge states of the 
dot which are almost degenerate in the vicinity of the Coulomb 
blockade peak. For $N_0^\past$ close to $N_0^*$ these are 
the states with $N = N_0^*\pm 1/2$ electrons on the dot. 
Hereafter we denote these states as $|N\rangle$ and 
$|N+1\rangle$. Finally, condition~\eref{3.2} enables 
us to replace the discrete single-particle levels within the 
dot by a continuum with the density of states $1/\delta E$.

Applying the Fermi Golden Rule and using the described 
simplifications, we may write the current $I_\alpha$ from 
the lead $\alpha$ into the dot as
\begin{eqnarray}
I_{\alpha} = \frac{2\pi}{\hbar}\sum_{kns} |t_{\alpha n}|^2 
\delta(\xi_k + eV_\alpha + E_N - \epsilon_n -E_{N+1}) 
\nonumber \\ 
~~~~~~~~~\times\left\{P_{N} f(\xi_k)[1-f(\epsilon_n)] 
- P_{N+1} f(\epsilon_n)[1-f(\xi_k)]\right\}.
\nonumber
\end{eqnarray}
Here $f(\omega) = [\exp(\omega/T)+1]^{-1}$ is the Fermi 
function, $V_\alpha$ is the potential on the lead $\alpha$, 
see Fig.~\ref{circuit}, $E_{N}$ and $E_{N+1}$  are the 
electrostatic energies of the charge states $|N\rangle$ and 
$|N+1\rangle$, and $P_N$ and $P_{N+1}$ are the 
probabilities to find the dot in these states. Replacing the 
summations over $n$ and $k$ by integrations over the 
corresponding continua, we find
\begin{equation}
I_\alpha = \frac{G_\alpha}{e}\left[P_N F(E_1-E_0- eV_\alpha)
- P_{N+1} F(E_0-E_1+ eV_\alpha)\right]
\label{3.4}
\end{equation}
with $F(\omega)=\omega/[\exp(\omega/T)-1]$. In equilibrium, 
the current~\eref{3.4} is zero by the detailed balance. When 
a finite current flows through the junction the probabilities 
$P_N$ and $P_{N+1}$ deviate from their equilibrium values. 
In the steady state, the currents across the two junctions satisfy
\begin{equation}
I = I_L = -I_R .
\label{3.5}
\end{equation}
Equations~\eref{3.4} and~\eref{3.5}, supplemented by the 
obvious normalization condition $P_N + P_{N+1} =1$, 
allow one to find $P_N$, $P_{N+1}$, and the current across 
the dot $I$ in response to the applied bias $V = V_L - V_R$.  
The linear conductance across the dot 
$G = \lim_{V\to 0}dI/dV$ is then given by~\cite{rate}
\begin{equation}
G = G_\infty
\frac{E_C (N_0^\past - N_0^*)/T}{\sinh [2E_C (N_0^\past - N_0^*)/T]} \,.
\label{3.6}
\end{equation}
Here $N_0^\past - N_0^* =0$ (half-integer $N_0$) corresponds 
to the Coulomb blockade peak. In the \textit{Coulomb blockade 
valleys} ($N_0^\past \neq N_0^*$), conductance falls off 
exponentially with the decrease of temperature, and all the valleys 
behave exactly the same way.

\section{Activationless transport through a blockaded quantum dot}

According to the rate equations theory~\cite{rate}, at low 
temperatures, $T\ll E_C $, conduction through the dot is 
exponentially suppressed in the Coulomb blockade valleys. 
This suppression occurs because the process of electron 
transport through the dot involves a \textit{real transition} 
to the state in which the charge of the dot differs by $e$ from 
the thermodynamically most probable value. The probability 
of such fluctuation is proportional to 
$\exp\left(-E_C |N_0^\past - N_0^*|/T\right)$, 
which explains the conductance suppression, see Eq.~\eref{3.5}. 
Going beyond the lowest-order perturbation theory in 
conductances $G_\alpha$ allows one to consider processes 
in which states of the dot with a ``wrong'' charge participate 
in the tunneling process as \textit{virtual states}. The existence 
of these higher-order contributions to the tunneling conductance 
was envisioned first by Giaever and Zeller~\cite{Giaever}. 
The first quantitative theory of this effect, however, was 
developed much later~\cite{AN}. 

The leading contributions to the activationless transport, 
according to~\cite{AN}, are provided by the processes 
of \textit{inelastic and elastic co-tunneling}. Unlike the sequential 
tunneling, in the co-tunneling mechanism, the events of electron 
tunneling from one of the leads into the dot, and tunneling from 
the dot to the other lead occur as a single quantum process. 

\subsection{Inelastic co-tunneling}

\begin{figure}[h]
\centerline{\includegraphics[width=0.95\textwidth]{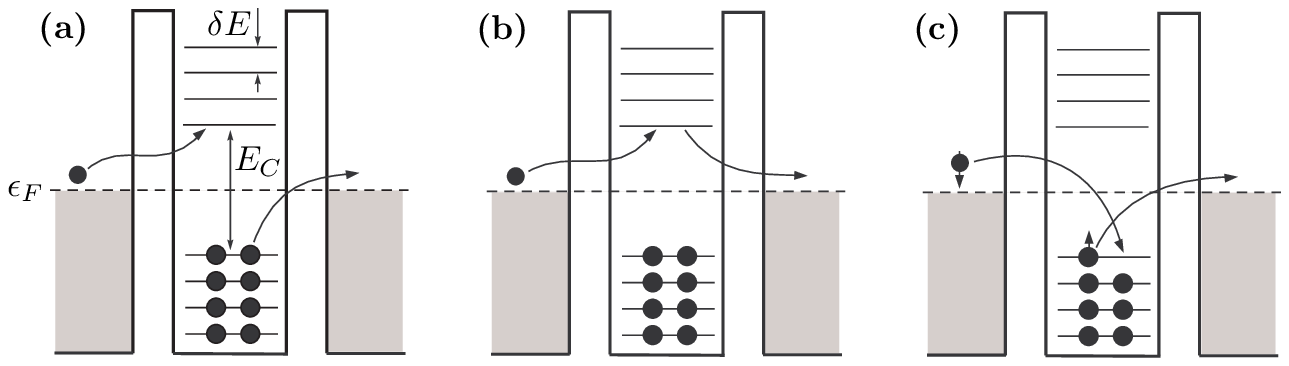}}
\vspace{2mm}
\caption{Examples of the co-tunneling processes. 
\newline (a) inelastic co-tunneling: transferring of an electron 
between the leads leaves behind an electron-hole pair in the dot;
(b) elastic co-tunneling;
(c) elastic co-tunneling with a flip of spin.
}
\label{cotunneling}
\end{figure}

In the inelastic co-tunneling mechanism, an electron tunnels 
from a lead into one of the vacant single-particle levels in 
the dot, while it is an electron occupying some other level 
that tunnels out of the dot, see Fig.~\ref{cotunneling}(a). As 
a result, transfer of charge $e$ between the leads is accompanied 
by a simultaneous creation of an electron-hole pair in the dot.

Here we will estimate the contribution of the inelastic co-tunneling 
to the conductance deep in the Coulomb blockade valley, i.e. 
at almost integer $N_0$. Consider an electron that tunnels into 
the dot from lead $L$. If energy $\omega$ of the electron 
relative to the Fermi level is small compared to the charging energy, 
$\omega\ll E_C $, then the energy of the virtual state involved in the
co-tunneling process is close to $E_C $. The amplitude $A_{in}$ 
of the inelastic transition via this virtual state to lead $R$ is then 
given by
\begin{equation}
A_{in} 
= \frac{t^\ast_{L n^\pprime} {\negthinspace} t^\past_{R n'}}{E_C }.
\label{4.1}
\end{equation}
The initial state of this transition has an extra electron in the
single-particle state $k$ in lead $L$, while the final state has an extra 
electron in the state $k'$ in lead $R$ and an electron-hole pair in the 
dot (state $n$ is occupied, state $n'$ is empty).  

Given the energy of the initial state $\omega$, the number of available 
final states can be estimated from the phase space argument, familiar 
from the calculation of the quasiparticle lifetime in the Fermi liquid 
theory~\cite{Abrikosov}. For $\omega\gg\delta E$ this number is of 
the order of $(\omega/\delta E)^2$. Since the typical value of $\omega$ 
is $T$, the inelastic co-tunneling contribution to the conductance can 
be estimated as 
\[
G_{in} \sim \frac{e^2}{h} \left(\frac{T}{\delta E}\right)^2 
 \nu^2 \overline{|A_{in}^2|} \,.
\]
Using now Equations \eref{2.17} and \eref{2.18}, we find~\cite{AN}
\begin{equation}
G_{in} \sim \frac{h}{e^2}\,G_LG_R\left(\frac{T}{E_C }\right)^2.
\label{4.2}
\end{equation}
A comparison of Eq.~\eref{4.2} with the result of the rate equations
theory (\ref{3.6}) shows that the inelastic co-tunneling takes over
the thermally-activated hopping at moderately low temperatures
\begin{equation}
T\lesssim T_{\rm in} =
E_C \left[\ln\left(\frac{e^2/h}{G_L+G_R}\right)\right]^{-1}.
\label{4.3}
\end{equation}

The smallest energy of the electron-hole pair is of the order of 
$\delta E$. At temperatures below this threshold the inelastic 
co-tunneling contribution to the conductance becomes exponentially 
small. It turns out, however, that even at much higher temperatures 
this mechanism becomes less effective than the elastic co-tunneling.

\subsection{Elastic co-tunneling}

In the process of elastic co-tunneling, transfer of charge between 
the leads is not accompanied by the creation of an electron-hole pair 
in the dot. In other words, occupation numbers of single-particle 
energy levels in the dot in the initial and final states of the co-tunneling 
process are exactly the same, see Fig.~\ref{cotunneling}(b). Close to 
the middle of the Coulomb blockade valley (at almost integer $N_0$)
the average number of electrons on the dot, $N\approx N_0$, is also 
an integer. Both an addition and a removal of a single electron cost 
$E_C$ in electrostatic energy, see Eq.~\eref{2.12}. The amplitude of the 
elastic co-tunneling process in which an electron is transfered from 
lead $L$ to lead $R$ can then be written as
\begin{equation}
A_{el} = \sum_n A_n,
\quad
A_n = t^\ast_{Ln} t^\past_{Rn}
\frac{\theta(\epsilon_n) - \theta(- \epsilon_n)}{E_C + |\epsilon_n|}\,.
\label{4.4}
\end{equation}
The two contributions to the partial amplitude $A_n$ are associated 
with virtual creation of either an electron if the level $n$ is empty 
($\epsilon_n > 0$), or of a hole if the level is occupied ($\epsilon_n < 0$); 
the relative sign difference between the two contributions originates 
in the fermionic commutation relations.

With the help of Equations~\eref{2.17} and \eref{2.18} the average 
value of the elastic co-tunneling contribution to the conductance can 
be written as
\[
\overline {G_{el}} = \frac{2e^2}{h}
\nu^2 \overline{\left| A^2_{el}\right|} 
\sim \frac{h}{e^2} \, G_L G_R 
\sum_n \left(\frac{\delta E}{E_C + |\epsilon_n|}\right)^2 .
\]
Since for $E_C\gg \delta E$ the number of terms making significant 
contribution to the sum over $n$ here is large, and since the sum is 
converging, one can replace the summation by an integral which 
results in~\cite{AN}
\begin{equation}
\overline{G_{el}}
\sim \frac{h}{e^2} \, G_L G_R  \frac{\delta E}{E_C }\,.
\label{4.5}
\end{equation}
Comparison with Eq.~\eref{4.2} shows that the elastic co-tunneling 
dominates the electron transport already at temperatures
\begin{equation}
T\lesssim T_{el} = \sqrt{E_C \delta E} \,,
\label{4.6}
\end{equation}
which may exceed significantly the level spacing. Note, however, that 
mesoscopic fluctuations of $G_{el}$ are strong~\cite{AG}, of the 
order of its average value \eref{4.5}. Thus, although $\overline{G_{el}}$ 
is always positive, see Eq.~\eref{4.6}, the sample-specific value of $G_{el}$ 
for a given gate voltage may vanish~\cite{Gefen}.  

\section{Effective low-energy Hamiltonian}
\label{H_K}

In the above discussion of the elastic co-tunneling we made a tacit 
assumption that all single-particle levels in the dot are either empty 
or doubly occupied. This, however, is not the case when the dot 
has a non-zero spin in the ground state. A dot with odd number 
of electrons, for example, would necessarily have a half-integer 
spin $S$. In the most important case of $S=1/2$ the top-most 
occupied single-particle level is filled by a single electron and is 
spin-degenerate. This opens a possibility of a co-tunneling process 
in which a transfer of an electron between the leads is accompanied 
by a flip of electron's spin with simultaneous flip of the spin on the 
dot, see Fig.~\ref{cotunneling}(c). 

The amplitude of such a process, calculated in the fourth order 
in tunneling matrix elements, diverges logarithmically when the 
energy $\omega$ of an incoming electron approaches $0$. Since 
$\omega\sim T$, the logarithmic singularity in the transmission 
amplitude translates into a dramatic enhancement of the 
conductance $G$ across the dot at low temperatures: $G$ may 
reach values as high as the quantum limit $2e^2/h$~\cite{AM, unitary}. 
This conductance enhancement is not really a surprise. Indeed, in 
the spin-flip co-tunneling process a quantum dot with odd $N$ 
behaves as $S=1/2$ magnetic impurity embedded into a tunneling 
barrier separating two massive conductors~\cite{Erice}. It is 
known~\cite{old_reviews} since mid-60's that the presence 
of such impurities leads to zero-bias anomalies in tunneling 
conductance~\cite{classics}, which are adequately 
explained~\cite{Appelbaum,Anderson} in the context of the 
Kondo effect~\cite{Kondo}. 

At energies well below the threshold $\Delta\sim\delta E$ for 
intradot excitations the transitions within the $(2S+1)$-fold 
degenerate ground state manifold of a dot can be conveniently 
described by a spin operator ${\bf S}$. The form of the 
\textit{effective Hamiltonian} describing the interaction of the 
dot with conduction electrons in the leads is then dictated by 
$SU(2)$ symmetry\footnote{In writing Eq.~\eref{5.1} we omitted the 
potential scattering terms associated with the usual elastic 
co-tunneling. This approximation is well justified when the 
conductances of the dot-lead junctions are small, 
$G_\alpha \ll e^2/h$, in which case $G_{el}$ is also very small, 
see Eq.~\eref{4.5}},
\begin{equation}
H_{\rm ef\mbox{}f} = \sum_{\alpha ks}\xi^\pdag_{k} 
c^\dagger_{\alpha ks} c^\pdag_{\alpha ks}
+ \sum_{\alpha \alpha'} J_{\alpha\alpha'}
({\bf s}_{\alpha'\alpha} \cdot {\bf S}) 
\label{5.1} 
\end{equation}
with ${\bf s}_{\alpha \alpha'} = \sum_{kk'ss'}
c_{\alpha ks}^{\dagger} (\bsigma_{ss'}/2) \,c_{\alpha' k's'}^\pdag$. 
The sum over $k$ in Eq.~\eref{5.1} is restricted to $|\xi_k|<\Delta$. 
The exchange amplitudes $J^\past_{\alpha\alpha'}$ form a
$2\times 2$ Hermitian matrix $\hat J$. The matrix has two 
real eigenvalues, the exchange constants $J_1$ and $J_2$ 
(hereafter we assume that $J_1\geq J_2$) By an appropriate 
rotation in the $R-L$ space the Hamiltonian~\eref{5.2} can 
then be brought into the form
\begin{equation}
H_{\rm ef\mbox{}f} = \sum_{\gamma ks} \xi^\pdag_{k} 
\psi^\dagger_{\gamma ks} \psi^\pdag_{\gamma ks}
+ \sum_{\gamma} J_\gamma
({\bf s}_\gamma\cdot{\bf S}).
\label{5.2}
\end{equation}
Here the operators $\psi_\gamma$ are certain linear combinations 
of the original operators $c_{R,L}$ describing electrons in the 
leads, and 
\[
{\bf s}_\gamma 
= \sum_{kk'ss'} \psi_{\gamma ks}^{\dagger} \frac{\bsigma_{ss'}}{2} 
\,\psi_{\gamma k's'}^\pdag
\] 
is local spin density of itinerant electrons in the ``channel" 
$\gamma =1,2$. 

The symmetry alone is not sufficient to determine the exchange 
constants $J_\gamma$;  their evaluation must rely upon a microscopic 
model. Here we breifly outline the derivation~\cite{real,Fiete} of 
Eq.~\eref{5.1} for a generic model of a quantum dot system discussed 
in Sec.~\ref{CIM} above. For simplicity, we will assume that 
the gate voltage $N_0$ is tuned to the middle of the Coulomb 
blockade valley.  The tunneling~\eref{2.16} mixes the state with 
$N = N_0$ electrons on the dot with states having $N\pm 1$ 
electrons. The electrostatic energies of these states are high 
$(\sim E_C )$, hence the transitions $N\to N\pm 1$ are virtual, 
and can be taken into account perturbatively in the second order 
in tunneling amplitudes~\cite{SW}. 

For Hamiltonian~\eref{2.12} the occupations of single-particle 
energy levels are good quantum numbers. Therefore, the amplitude 
$J_{\alpha\alpha'}$ can be written as a sum of partial amplitudes,
\begin{equation}
J_{\alpha\alpha'} = \sum_n J_{\alpha\alpha'}^n .
\label{5.3}
\end{equation}
Each term in the sum here corresponds to a process during which 
an electron or a hole is created virtually on level $n$ in the dot, 
cf. Eq.~\eref{4.4}. For $G_\alpha \ll e^2/h$ and $E_S \ll \delta E$ the 
main contribution to the sum in~\eref{5.3} comes from 
singly-occupied energy levels in the dot. A dot with spin $S$ has 
$2S$ such levels near the Fermi level (hereafter we assign indexes 
$n= -S,\ldots,n=S$ to these levels), each carrying a spin 
${\bf S}/2S$, and contributing 
\begin{equation}
J_{\alpha\alpha'}^n = \frac{\lambda_n}{E_C}\,
t^\ast_{\alpha n}  t^\past_{\alpha' n},
\quad 
\lambda_n = 2/S,
\quad
|n|\leq S
\label{5.4}
\end{equation}
to the exchange amplitude in~\eref{5.1}. This yields
\begin{equation}
J_{\alpha\alpha'}  
\approx \sum_{|n|\leq S} J_{\alpha\alpha'}^n .
\label{5.5}
\end{equation}
It follows from Equations~\eref{5.3} and~\eref{5.4} that 
\begin{equation}
\tr \hat J = \frac{1}{E_C} \sum_n \lambda_n 
\left(|t^2_{Ln}| + |t^2_{Rn}|\right) .
\label{5.6}
\end{equation}
By restricting the sum over $n$ here to $|n|\leq S$, as in~\eref{5.5}, 
and taking into account that all $\lambda_n$ in~\eref{5.4} are positive, 
we find $J_1 + J_2 > 0$. Similarly, from
\begin{equation}
\det\hat J = \frac{1}{2E^2_C} \sum_{m,n} \lambda_m \lambda_n 
|\mathcal D_{mn}^2|,
\quad
\mathcal D_{mn} = \det 
\left(
\begin{array}{lr}
t_{Lm}& t_{Rm} 
\\
t_{Ln}& t_{Rn} 
\end{array}
\right)
\label{5.7}
\end{equation}
and Equations~\eref{5.4} and~\eref{5.5} it follows that $J_1 J_2 > 0$ 
for $S>1/2$. Indeed, in this case the sum in~\eref{5.7} contains at least 
one contribution with $m\neq n$; all such contributions are positive.
Thus, both exchange constants $J_{1,2} > 0$ if the dot's spin 
$S$ exceeds $1/2$~\cite{real}. The pecularities of the Kondo effect 
in quantum dots with large spin are discussed in Sec.~\ref{large_S} 
below. 

Here we concentrate on the most common situation of $S=1/2$ on 
the dot~\cite{kondo_exp}, considered in detail in Sec.~\ref{S_half}. 
The ground state of such dot has only one singly-occupied energy level 
$(n=0)$, so that $\det\hat J \approx 0$, see~\eref{5.5} and \eref{5.7}.  
Accordingly, one of the exchange constants vanishes,
\begin{equation}
J_2 \approx 0,
\label{5.8}
\end{equation}
while the remaining one, $J_1 = \tr\hat J$, is positive. \Eref{5.8} 
resulted, of course, from the approximation made in~\eref{5.5}. 
For the model~\eref{2.12} the leading correction to~\eref{5.5} 
originates in the co-tunneling processes with an intermediate state 
containing an extra electron (or an extra hole) on one of the empty 
(doubly-occupied) levels. Such contribution arises because the spin 
on the level $n$ is not conserved by the Hamiltonian~\eref{2.12}, 
unlike the corresponding occupation number. Straightforward 
calculation~\cite{Fiete} yields the partial amplitude in the form 
of \mbox{Eq.~\eref{5.4}}, but with
\[
\lambda_n = - \frac{2E_C E_S}{(E_C + |\epsilon_n|)^2},
\quad
\quad n\neq 0 .
\]

Unless the tunneling amplitudes $t_{\alpha 0}$ to the only 
singly-occupied level in the dot are anomalously small, the 
corresponding correction 
\begin{equation}
\delta J_{\alpha\alpha'} = \sum_{n\neq 0} J_{\alpha\alpha'}^n 
\label{5.9}
\end{equation}
to the exchange amplitude~\eref{5.5} is small,
\[
\left|\frac{\delta J_{\alpha\alpha'}}
{J_{\alpha\alpha'}}\right| \sim \frac{E_S}{\delta E} \ll 1,
\]
see Eq.~\eref{2.13}. To obtain this estimate, we assumed that all 
tunneling amplitudes $t_{\alpha n}$ are of the same order of 
magnitude, and replaced the sum over $n$ in~\eref{5.9} by 
an integral. A similar estimate yields the leading contribution 
to $\det\hat J$,
\[
\det \hat J \approx \frac{1}{E^2_C} \sum_n \lambda_0 \lambda_n 
|\mathcal D_{0n}^2|
\sim - \frac{E_S}{\delta E} \bigl(\tr\hat J\bigr)^2,
\]
or
\begin{equation}
J_2/J_1 \sim - E_S/\delta E .
\label{5.10}
\end{equation}

According to~\eref{5.10}, the exchange constant $J_2$ is
negative~\cite{SI}, and its absolute value is small compared to $J_1$.
Hence, \eref{5.8} is indeed an excellent approximation for large
chaotic dots with spin $S=1/2$ as long as the intradot exchange
interaction remains weak, $E_S \ll \delta E$\footnote{\Eref{5.8} holds
identically for the Anderson impurity model~\cite{Anderson}
frequently employed to study transport through quantum 
dots~\cite{AM,MWL}. In that model a quantum dot is described by 
a single energy level, which formally corresponds to the infinite
level spacing limit $\delta E\to\infty$ of the Hamiltonian~\eref{2.12}.}.  
Note that corrections to the universal Hamiltonian~\eref{2.12} also 
result in finite values of both exchange constants, $|J_2|\sim J_1 N^{-1/2}$, 
and become important for small dots with $N\lesssim 10$~\cite{unitary}. 
Although this may significantly affect the conductance across the system 
in the weak coupling regime $T\gtrsim T_K$, it does not lead to qualitative 
changes in the results for $S = 1/2$ on the dot, as the channel with 
smaller exchange constant decouples at low energies~\cite{NB}, see 
also Sec.~\ref{large_S} below. With this caveat, we adopt  
approximation~\eref{5.8} in our description of the Kondo effect in 
quantum dots with spin $S=1/2$. Accordingly, the effective Hamiltonian 
of the system~\eref{5.2} assumes the``block-diagonal'' form
\begin{eqnarray}
H_{\rm ef\mbox{}f} = H_1 + H_2 
\label{5.11} \\
H_1 = \sum_{ks}\xi^\pdag_{k} \psi^\dagger_{1 ks} \psi^\pdag_{1 ks}
+ J ({\bf s}_1 \cdot {\bf S}) 
\label{5.12} \\
H_2 = \sum_{ks}\xi^\pdag_{k} \psi^\dagger_{2ks} \psi^\pdag_{2ks}
\label{5.13} 
\end{eqnarray}
with $J = \tr \hat J > 0$. 

\section{Kondo regime in transport through a quantum dot}
\label{S_half}

To get an idea about the physics of the Kondo model 
(see~\cite{Kondo_reviews} for recent reviews), let us first 
replace the fermion field operator ${\bf s}_1$ in Eq.~\eref{5.12} 
by a single-particle spin-1/2 operator ${\bf S}_1$. The ground 
state of the resulting Hamiltonian of two spins 
\[
\widetilde{H} = J({\bf S}_1\cdot{\bf S})
\] 
is obviously a singlet. The excited state (a triplet) is separated 
from the ground state by the energy gap $J_1$. This separation 
can be interpreted as the binding energy of the singlet. Unlike 
${\bf S}_1$ in this simple example, the operator ${\bf s}_1$ 
in~\eref{5.12} is merely a spin density of the conduction electrons 
at the site of the ``magnetic impurity". Because conduction electrons 
are freely moving in space, it is hard for the impurity to ``capture" 
an electron and form a singlet. Yet, even a weak local exchange 
interaction suffices to form a singlet ground 
state~\cite{PWA_book,Wilson}. 
However, the characteristic energy (an analogue of the binding 
energy) for this singlet is given not by the exchange constant $J$, 
but by the so-called Kondo temperature
\begin{equation}
T_K \sim \Delta \exp(-1/\nu J).
\label{6.1}
\end{equation}
Using $\Delta\sim\delta E$ and Equations~\eref{5.6} and~\eref{2.18}, 
one obtains from~\eref{6.1} the estimate
\begin{equation}
\ln\left(\frac{\delta E}{T_K}\right)
\sim \frac{1}{\nu J}
\sim \frac{e^2/h}{G_L + G_R}\frac{E_C }{\delta E}.
\label{6.2}
\end{equation}
Since $G_\alpha\ll e^2/h$ and $E_C \gg \delta E$, the r.h.s. 
of~\eref{6.2} is a product of two large parameters. Therefore, 
the Kondo temperature $T_K $ is small compared to the mean 
level spacing,
\begin{equation}
T_K \ll \delta E.
\label{6.3}
\end{equation}
It is this separation of the energy scales that justifies the use of the 
effective low-energy Hamiltonian \eref{5.1},~\eref{5.2} for the 
description of the Kondo effect in a quantum dot system. 
Inequality~\eref{6.3} remains valid even if the conductances 
of the dot-leads junctions $G_\alpha$ are of the order of $2e^2/h$. 
However, in this case the estimate~\eref{6.2} is no longer
applicable~\cite{GHL}. 

In our model, see Equations~\eref{5.11}-\eref{5.13}, one of 
the channels $(\psi_2)$ of conduction electrons completely 
decouples from the dot, while the $\psi_1$-particles are described 
by the standard single-channel antiferromagnetic Kondo 
model~\cite{Kondo,Kondo_reviews}. Therefore, the 
thermodynamic properties of a quantum dot in the Kondo regime 
are identical to those of the conventional Kondo problem for a 
single magnetic impurity in a bulk metal; thermodynamics of the 
latter model is fully studied~\cite{bethe}.  However, all the 
experiments addressing the Kondo effect in quantum dots test 
their transport properties rather then thermodynamics. The electron 
current operator is not diagonal in the $(\psi_1,\psi_2)$ representation, 
and the contributions of these two sub-systems to the conductance 
are not additive. Below we relate the linear conductance and, in some 
special case, the non-linear differential conductance as well, to the 
t-matrix of the conventional Kondo problem.

\subsection{Linear response}

The linear conductance can be calculated from the Kubo formula
\begin{equation}
G = \lim_{\omega\to 0} \frac{1}{\hbar\,\omega}
\int_0^\infty d t \,\rme^{i\omega t} 
\Bigl\langle \bigl[\hat I(t), \hat I(0) \bigr]\Bigr\rangle ,
\label{6.4} 
\end{equation}
where the current operator $\hat I$ is given by
\begin{equation}
\hat I = \frac{d}{d t} \frac{e}{2} 
\bigl(\hat N_{R} - \hat N_{L}\bigr), 
\quad
\hat N_\alpha 
= \sum_{ks}^\pdag c^\dagger_{\alpha ks}c^\pdag_{\alpha ks}
\label{6.5} 
\end{equation}
Here $\hat N_\alpha$ is the operator of the total number of electrons 
in the lead $\alpha$. Evaluation of the linear conductance proceeds 
similarly to the calculation of the impurity contribution to the resistivity 
of dilute magnetic alloys (see, e.g.,~\cite{AL}). In order to take the 
full advantage of the decomposition~\eref{5.11}-\eref{5.13}, we rewrite 
$\hat I$ in terms of the operators $\psi_{1,2}$. These operators are 
related to the original operators $c_{R,L}$ representing the electrons 
in the right and left leads via
\begin{equation}
\left(\begin{array}{c}
\psi_{1ks} \\ 
\psi_{2ks}
\end{array}\right)
= \left(\begin{array}{cc}
\cos\theta_0 & \sin\theta_0
\\ 
-\sin\theta_0 & \cos\theta_0 
\end{array}\right)
\left(\begin{array}{c}
c_{Rks} \\ 
c_{Lks}
\end{array}\right) .
\label{6.6} 
\end{equation}
The rotation matrix here is the same one that diagonalizes matrix
$\hat J$ of the exchange amplitudes in~\eref{5.1}; the rotation angle
$\theta_0$ satisfies the equation $\tan \theta_0 = |t_{L0}/t_{R0}|$.
With the help of Eq.~\eref{6.6} we obtain
\begin{equation}
\hat N_R - \hat N_L = \cos(2\theta_0) \bigl(\hat N_1 - \hat N_2\bigr) 
- \sin(2\theta_0) \sum_{ks}
\bigl(
\psi_{1ks}^\dagger \psi_{2ks}^\pdag + 
{\rm H.c.}
\bigr) 
\label{6.7} 
\end{equation}
The current operator $\hat I$ entering the Kubo formula~\eref{6.4} 
is to be calculated with the equilibrium Hamiltonian~\eref{5.11}-\eref{5.13}. 
Since both $\hat N_1$ and $\hat N_2$ commute with $H_{\rm ef\mbox{}f}$, 
the first term in~\eref{6.7} makes no contribution to $\hat I$. When 
the second term in~\eref{6.7} is substituted into~\eref{6.5} and then 
into the Kubo formula~\eref{6.4}, the result, after integration by parts, 
can be expressed via 2-particle correlation functions such as 
$\bigl\langle \psi^\dagger_1(t)\psi_2^\pdag(t) 
\psi_2^\dagger(0)\psi_1^\pdag(0)\bigr\rangle$ 
(see Appendix B of~\cite{PG} for further details about this calculation). 
Due to the block-diagonal structure of $H_{\rm ef\mbox{}f}$, 
see~\eref{5.11}, these correlation functions factorize into products 
of the single-particle correlation functions describing the (free) 
$\psi_2$-particles and the (interacting) $\psi_1$-particles. The result 
of the evaluation of the Kubo formula can then be written as 
\begin{equation}
G = G_0 \int d\omega 
\left(-\frac{d f}{d \omega}\right) 
\frac{1}{2} \sum_s \bigl[- \pi\nu \,{\rm Im}\, T_s (\omega)\bigr]. 
\label{6.8}
\end{equation}
Here 
\begin{equation}
G_0 = \frac{2e^2}{h}  \sin^2(2\theta_0) 
= \frac{2e^2}{h} 
\frac{4|t^2_{L0} t^2_{R0}|}{\bigl(|t_{L0}^2|+|t_{R0}^2|\bigr)^2} \,,
\label{6.9}
\end{equation}
\noindent 
$f(\omega)$ is the Fermi function, and $T_s(\omega)$ is the 
t-matrix for the Kondo model~\eref{5.12}. The t-matrix is 
related to the exact retarded Green function of the 
$\psi_1$-particles in the conventional way,
\[
G_{ks,k'\negthinspace s}(\omega) 
= G^0_k(\omega) + G^0_k (\omega)T_s (\omega) G^0_{k'}(\omega), 
\quad
G^0_k = (\omega - \xi_k +i0)^{-1} .
\]
Here $G_{ks,k'\negthinspace s}(\omega)$ is the Fourier transform of 
$G_{ks,k'\negthinspace s}(t) = -i\theta(t)
\bigl\langle\bigl\{
\psi_{1ks}^\pdag(t),\psi_{1k'\negthinspace s}^\dagger
\bigr\}\bigr\rangle$, 
where $\langle\ldots\rangle$ stands for the thermodynamic averaging 
with Hamiltonian~\eref{5.12}. In writing Eq.~\eref{6.8} we took into 
account the conservation of the total spin (which implies that 
$G_{ks,k'\negthinspace s'}=\delta_{ss'}G_{ks,k'\negthinspace s}$, 
and that the interaction in~\eref{5.12} is local (which in turn means 
that the t-matrix is independent of $k$ and $k'$). 

\subsection{Weak coupling regime: $T_K\ll T\ll \delta E$}

When the exchange term in the Hamiltonian~\eref{5.12} is treated 
perturbatively, the main contribution to the t-matrix comes from 
the transitions of the type~\cite{AAA}
\begin{equation}
\ket{ks,\sigma}\to\ket{k'\negthinspace s'\negthinspace ,\sigma'}.
\label{1}
\end{equation}
Here state $\ket{ks,\sigma}$ has an extra electron with spin $s$ in 
the orbital state $k$ whereas the dot is in the spin state $\sigma$. 
By $SU(2)$ symmetry, the amplitude of the transition~\eref{1}
satisfies
\begin{equation}
A_{\ket{k'\negthinspace s'\negthinspace ,\sigma'}\leftarrow \ket{ks,\sigma}} 
= A_{k^\prime k^\pprime} \frac{1}{4} 
\left(\bsigma_{s'\negthinspace s}\cdot\bsigma_{\sigma'\negthinspace \sigma}\right)
\label{2}
\end{equation}
The transition~\eref{1} is \textit{elastic} in the sense that the
number of quasiparticles in the final state of the transition is the
same as that in the initial state (in other words, the transition~\eref{1} 
is not accompanied by the production of electron-hole pairs). Therefore, 
the imaginary part of the t-matrix can be calculated with the help of the 
optical theorem~\cite{Newton}, which yields
\begin{equation}
-\pi\nu \,{\rm Im}\, T_s 
= \frac{1}{2}\sum_\sigma \sum_{s'\negthinspace\sigma'}
\left| \pi\nu \,
A_{\ket{k'\negthinspace s'\negthinspace ,\sigma'}
\leftarrow \ket{ks,\sigma}}^2
\right|\,.
\label{3}
\end{equation}
The factor $1/2$ here accounts for the probability to have spin 
$\sigma$ on the dot in the initial state of the transition. Substitution 
of the tunneling amplitude in the form~\eref{2} into Eq.~\eref{3}, and 
summation over the spin indexes with the help of the identity~\eref{0} 
result in
\begin{equation}
-\pi\nu \,{\rm Im}\, T_s = \frac{3\pi^2}{16\,} \,\nu^2 \left|A_{k'k}^2\right| .
\label{4}
\end{equation}

The amplitude $A_{k'\negthinspace k}$ in Equations~\eref{2} and~\eref{4} 
depends only on the difference of energies $\omega = \xi_{k'} - \xi_k$,
\[
A_{k'\negthinspace k} = A(\omega).
\] 
In the leading (first) order in $J$ one readily obtains $A^{(1)} = J$, 
independently of $\omega$. However, as discovered by Kondo~\cite{Kondo},
the second-order contribution $A^{(2)}$ not only depends on $\omega$,
but is logarithmically divergent as $\omega\to 0$:
\[
A^{(2)}(\omega) = \nu J^2 \ln\left|\Delta/\omega\right|. 
\]
Here $\Delta$ is the high-energy cutoff in the Hamiltonian~\eref{5.12}.
It turns out~\cite{AAA} that similar logarithmically divergent contributions
appear in all orders of perturbation theory, 
\[
\nu A^{(n)} (\omega) = (\nu J)^n \bigl[\ln|\Delta/\omega|\bigr]^{n-1},
\]
resulting in a geometric series
\[
\nu A(\omega) = \sum_{n=1}^{\infty} \nu A^{(n)} 
= \nu J \sum_{n=0}^{\infty}\bigl[\nu J \ln|\Delta/\omega|\bigr]^n
=  \frac{\nu J}{1 - \nu J\ln|\Delta/\omega|} \,.
\]
With the help of the definition of the Kondo temperature~\eref{6.1}, 
this can be written as
\begin{equation}
\nu A(\omega) = \frac{1}{\ln|\omega/T_K|}\,.
\label{5}
\end{equation}
Substitution of~\eref{5} into Eq.~\eref{4} and then into Eq.~\eref{6.8}, 
and evaluation of the integral over $\omega$ with logarithmic 
accuracy yield for the conductance across the dot
\begin{equation}
G = G_0 \frac{3\pi^2/16}{\ln^2(T/T_K)}\,, 
\quad
T\gg T_K .
\label{6.12}
\end{equation}
\Eref{6.12} is the leading term of the asymptotic expansion in 
powers of~$1/\ln(T/T_K)$, and represents the conductance 
in the {\it leading logarithmic approximation}. 

\Eref{6.12} resulted from summing up the most-diverging 
contributions in all orders of perturbation theory. It is instructive 
to re-derive it now in the framework of
\textit{renormalization group}~\cite{PWA}. The idea of this 
approach rests on the observation that the electronic states that
give a significant contribution to observable quantities, such as 
conductance, are states within an interval of energies of the 
width $\omega\sim T$ about the Fermi level, see Eq.~\eref{6.8}. 
At temperatures of the order of $T_K$, when the Kondo effect 
becomes important, this interval is narrow compared to the width
of the band $D=\Delta$ in which the Hamiltonian~\eref{5.12} is 
defined.

Consider a narrow strip of energies of the width $\delta D\ll D$ near 
the edge of the band. Any transition~\eref{1} between a state near the 
Fermi level and one of the states in the strip is associated with high 
($\sim\Delta$) energy deficit, and, therefore, can only occur virtually.
 Obviously, virtual transitions via each of the states in the strip result 
in the second-order correction $\sim J^2/D$ to the amplitude $A(\omega)$ 
of the transition between states in the vicinity of the Fermi level. Since 
the strip contains $\nu \delta D$ electronic states, the total correction 
is~\cite{PWA} 
\[
\delta A \sim \nu J^2\delta D/D .
\] 
This correction can be accounted for by modifying the exchange
constant in the effective Hamiltonian $\widetilde{H}_{\rm ef\mbox{}f}$
which is defined for states within a narrower energy band of the width
$D-\delta D$~\cite{PWA},
\begin{eqnarray}
\widetilde{H}_{\rm ef\mbox{}f}
= \sum_{ks} \xi^\pdag_k 
{\psi}^\dagger_{ks}{\psi}^\pdag_{ks}
+ J_{D - \delta D}
({\bf s}_\psi \cdot{\bf S}),
\quad
|\xi_k|<D-\delta D,
\label{6} \\
J_{D - \delta D} = J_D + \nu J_D^2 \frac{\delta D}{D}.
\label{7}
\end{eqnarray}
Here $J_D$ is the exchange constant in the original Hamiltonian. 
Note that the $\widetilde{H}_{\rm ef\mbox{}f}$ has the same 
form as Eq.~\eref{5.12}. This is not merely a conjecture, but can be 
shown rigorously~\cite{Wilson,AYH}. 

The reduction of the bandwidth can be considered to be a result 
of a unitary transformation that decouples the states near the band 
edges from the rest of the band. In principle, any such transformation 
should also affect the operators that describe the observable quantities. 
Fortunately, this is not the case for the problem at hand. Indeed, 
the angle $\theta_0$ in Eq.~\eref{6.6} is not modified by the transformation. 
Therefore, the current operator and the expression for the 
conductance~\eref{6.8} retain their form.

Successive reductions of $D$ by small steps $\delta D$ can be viewed 
as a continuous process during which the initial Hamiltonian~\eref{5.12} 
with $D=\Delta$ is transformed to an effective Hamiltonian of the same 
form that acts within the band of the reduced width $D\ll \Delta $. It 
follows from~\eref{7} that the dependence of the effective exchange 
constant on $D$ is described by the differential equation~\cite{PWA,AYH}
\begin{equation}
\frac{d J_D}{d\zeta} =\nu J_D^2,
\quad
\zeta = \ln\left({\Delta}/{D}\right).
\label{8}
\end{equation}
With the help of Eq.~\eref{6.1}, the solution of the RG equation~\eref{8} 
subject to the initial condition $J_\Delta = J$ can be cast into the form
\[
\nu J_D = \frac{1}{\ln(D/T_K)}\,.
\]
The renormalization described by Eq.~\eref{8} can be continued until 
the bandwidth $D$ becomes of the order of the typical energy 
$|\omega|\sim T$ for real transitions. After this limit has been reached, 
the transition amplitude $A(\omega)$ is calculated in lowest (first) 
order of perturbation theory in the effective exchange constant (higher 
order contributions are negligibly small for $D\sim \omega$ ),
\[
\nu A(\omega) = \nu J_{D\sim|\omega|} = \frac{1}{\ln|\omega/T_K|}
\]
Using now Equations~\eref{4} and~\eref{6.8}, we recover Eq.~\eref{6.12}.

\subsection{Strong coupling regime: $T\ll T_K$}

As temperature approaches $T_K$, the leading logarithmic 
approximation result~\eref{6.12} diverges. This divergence signals 
the failure of the approximation. Indeed, we are considering a model 
with single-mode junctions between the dot and the leads. The 
maximal possible conductance in this case is $2e^2/h$. To obtain 
a more precise bound, we discuss in this section the conductance 
in the strong coupling regime $T\ll T_K$.

We start with the zero-temperature limit $T=0$. As discussed 
above, the ground state of the Kondo model~\eref{5.12} is a 
singlet~\cite{PWA_book}, and, accordingly, is not degenerate. 
Therefore, the t-matrix of the conduction electrons interacting with 
the localized spin is completely characterized by the scattering phase 
shifts $\delta_s$ for electrons with spin $s$ at the Fermi level. The 
t-matrix is then given by the standard scattering theory 
expression~\cite{Newton}
\begin{equation}
-\pi\nu \,T_s(0) = \frac{1}{2i}\left(\mathbb{S}_s -1\right) ,
\quad \mathbb{S}_s = \rme^{2i\delta_s},
\label{6.13}
\end{equation}
where $\mathbb{S}_s$ is the scattering matrix for electrons with 
spin $s$, which for a single channel case reduces to its eignevalue. 
Substitution of~\eref{6.13} into Eq.~\eref{6.8} yields
\begin{equation}
G(0) = G_0\frac{1}{2} \sum_s \sin^2\delta_s
\label{6.14}
\end{equation}
for the conductance, see Eq.~\eref{6.8}. The phase shifts
in~\eref{6.13},~\eref{6.14} are obviously defined only ${\rm mod}\,\pi$
(that is, $\delta_s$ and $\delta_s +\pi$ are equivalent). This
ambiguity can be removed if we set to zero the values of the phase
shifts at $J = 0$ in Eq.~\eref{5.12}.

In order to find the two phase shifts $\delta_s$, we need two 
independent relations. The first one follows from the invariance 
of the Kondo Hamiltonian~(\ref{5.12}) under the particle-hole 
transformation $\psi^\pdag_{ks} \to s\psi^\dagger_{-k,-s}$ (here 
$s=\pm 1$ for spin-up/down electrons). The particle-hole symmetry 
implies the relation for the t-matrix
\begin{equation}
T_s(\omega) = - T^*_{-s}(-\omega),
\label{6.15}
\end{equation}valid at all $\omega$ and $T$. In view of 
Eq.~\eref{6.13}, it translates into the relation for the phase shifts 
at the Fermi level ($\omega = 0$)~\cite{N},
\begin{equation}
\delta_\uparrow + \delta_\downarrow = 0.
\label{6.16}
\end{equation}

The second relation follows from the requirement that the 
ground state of the Hamiltonian~\eref{5.12} is a singlet~\cite{N}. 
In the absence of exchange ($J=0$) and at $T=0$, an 
infinitesimally weak $(B\to + 0)$ magnetic field acting on
the dot's spin, 
\begin{equation}
H_B = BS^z,
\label{6.17}
\end{equation} 
would polarize it; here $B$ is the Zeeman energy. Since free 
electron gas has zero spin in the ground state, the total spin in a 
very large but finite region of space $\cal V$ surrounding the dot 
coincides with the spin of the dot, $\langle S^z\rangle_{J=0} = -1/2$. 
If the exchange with electron gas is now turned on, $J > 0$, a 
very weak field will not prevent the formation of a singlet 
ground state. In this state, the total spin within $\cal V$ is zero. Such 
change of the spin is possible if the numbers of spin-up and spin-down 
electrons within $\cal V$ have changed to compensate for the 
dot's spin: $\delta N_\uparrow -\delta N_\downarrow = 1$.
By the Friedel sum rule, $\delta N_s$ are related to the scattering phase 
shifts at the Fermi level, $\delta N_s = \delta_s/\pi $, which gives
\begin{equation}
\delta_\uparrow - \delta_\downarrow = \pi.
\label{6.18}
\end{equation}

Combining~\eref{6.16} and~\eref{6.18}, we find $|\delta_s| = \pi/2$.
\Eref{6.14} then yields for zero-temperature and zero-field
conductance across the dot~\cite{AM}
\begin{equation}
G(0) = G_0.
\label{6.19}
\end{equation}
Thus, the grow of the conductance with lowering the temperature 
is limited only by the value of $G_0$. This value, see Eq.~\eref{6.9}, 
depends only on the ratio of the tunneling amplitudes $|t_{L0}/t_{R0}|$.  
If $|t_{L0}|=|t_{R0}|$, then the conductance at $T=0$ will reach 
the maximal value $G=2e^2/h$ allowed by quantum mechanics~\cite{AM}.

\begin{figure}[h]
\centerline{\includegraphics[width=0.8\textwidth]{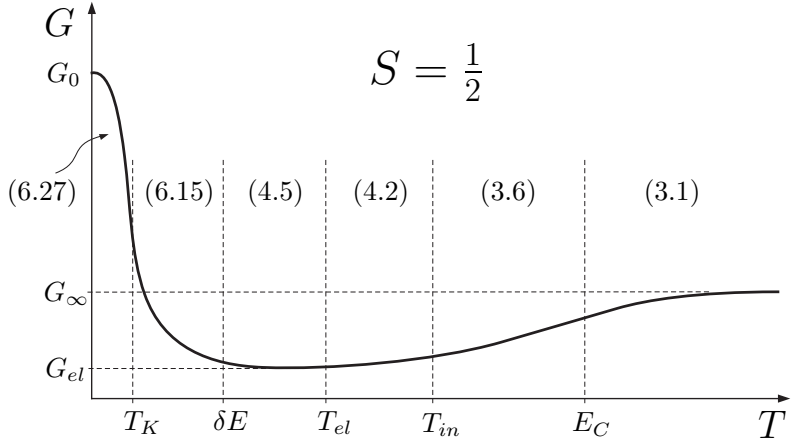}}
\vspace{2mm}
\caption{Sketch of the temperature dependence of the conductance 
in the middle of the Coulomb blockade valley with $S = 1/2$ on the 
dot. The numbers in brackets refer to the corresponding equations 
in the text.
}
\label{overall}
\end{figure}

The maximal conductance, Eq.~\eref{6.19}, is reached when a
a singlet state is formed by the itinerant electrons interacting with
the local spin, as described by the Kondo Hamiltonian~\eref{5.12}.
Perturbation of this singlet~\cite{N} by a magnetic field $B$ or
temperature $T$ leads to a decrease of the conductance. This decrease
is small as long as $B$ and $T$ are small compared to the singlet
``binding energy'' $T_K$. The reader is referred to the original
papers~\cite{N} for the details. Here we only
quote the result~\cite{AL} for the imaginary part of the t-matrix at
$|\omega|$ and $T$ small compared to the Kondo temperature $T_K$,
\begin{equation}
-\pi\nu \,{\rm Im}\, T_s (\omega) 
= 1-\frac{3\omega^2 +\pi^2T^2}{2\,T_K^2}\,.
\label{6.20}
\end{equation}
Substitution of~\eref{6.20} into~\eref{6.8} yields
\begin{equation}
G = G_0\left[1 - \left({\pi T}/{T_K}\right)^2\right],
\quad
T\ll T_K\,.
\label{6.21}
\end{equation}
Accordingly, corrections to the conductance are quadratic in 
temperature -- a typical Fermi liquid result~\cite{N}. The 
weak-coupling ($T\gg T_K$) and the strong-coupling ($T\ll T_K$)
asymptotes of the conductance have a strikingly different structure.
Nevertheless, since the Kondo effect is a crossover phenomenon 
rather than a phase transition~\cite{Kondo_reviews,PWA_book,Wilson,bethe}, 
the dependence $G(T)$ is a smooth and featureless~\cite{Costi} 
function throughout the crossover region $T\sim T_K$.

Finally, note that both Eq.~\eref{6.12} and Eq.~\eref{6.21} have
been obtained here for the particle-hole symmetric model~(\ref{5.12}). 
This approximation is equivalent to neglecting the elastic co-tunneling 
contribution to the conductance $G_{el}$. The asymptotes~\eref{6.12} 
and~\eref{6.21} remain valid~\cite{real} as long as $G_{el}/ G_0\ll 1$. 
The overall temperature dependence of the linear conductance in the 
middle of the Coulomb blockade valley is sketched in Fig.~\ref{overall}.

\subsection{Beyond linear response}
\label{NONEQ}

In order to study transport through a quantum dot away from 
equilibrium we add to the effective Hamiltonian~\eref{5.11}-\eref{5.13} 
a term 
\begin{equation}
H_V = \frac{eV}{2} \left(\hat N_L - \hat N_R\right)
\label{6.22}
\end{equation}
describing a finite voltage bias $V$ applied between the left and right 
electrodes. Here we will evaluate the current across the dot at arbitrary 
$V$ but under the simplifying assumption that the dot-lead junctions are 
strongly asymmetric: 
\[
G_L\ll G_R.
\] 
Under this condition the angle $\theta_0$ in~\eref{6.6} is small, 
$\theta_0\approx |t_{L0}/t_{R0}|\ll 1$. Expanding Eq.~\eref{6.7} 
to linear order in $\theta_0$ we obtain
\begin{equation}
H_V(\theta_0) = \frac{eV}{2} \bigl(\hat N_2- \hat N_1\bigr) 
+ eV \theta_0 \sum_{ks}
\bigl( \psi_{1ks}^\dagger \psi_{2ks}^\pdag + {\rm H.c.} \bigr)
\label{6.23} 
\end{equation}
The first term in the r.h.s. here can be interpreted as the voltage 
bias between the reservoirs of $1$- and $2$-particles, cf. Eq.~\eref{6.22}, 
while the second term has an appearance of $k$-conserving tunneling 
with very small (proportional to $\theta_0\ll 1$) tunneling amplitude. 

Similar to Eq.~\eref{6.23}, the current operator $\hat I$, see~\eref{6.5}, 
splits naturally into two parts,
\begin{eqnarray*}
\hat I = \hat I_0 + \delta \hat I, 
\\
 \hat I_0 = \frac{d}{d t} \frac{e}{2} \bigl(\hat N_1- \hat N_2\bigr) 
= - i e^2 V \theta_0 \sum_{ks} 
\psi_{1ks}^\dagger \psi_{2ks}^\pdag + {\rm H.c.} ,
\\
\delta\hat I = - e\theta_0 \frac{d}{d t} \sum_{ks} 
\psi_{1ks}^\dagger \psi_{2ks}^\pdag + {\rm H.c.}
\end{eqnarray*}
It turns out that $\delta\hat I$ does not contribute to the average 
current in the leading (second) order in $\theta_0$~\cite{Erice}. 
The remaining contribution $I = \bigl\langle\hat I_0\bigr\rangle$ 
corresponds to tunneling current between two bulk reservoirs 
containing $1$- and $2$-particles. Its evaluation yields~\cite{Erice}
\begin{equation}
\frac{d I}{d V} = G_0 \frac{1}{2} \sum_{s}
\left[- \pi\nu \,{\rm Im}\, T_s (eV)\right] 
\label{6.24}
\end{equation}
for the differential conductance across the dot at zero temperature. 
Here $G_0$ coincides with the small $\theta_0$-limit of Eq.~\eref{6.9}. 
Using now Equations~\eref{4}, \eref{5}, and~\eref{6.20}, we obtain 
\begin{equation}
\frac{1}{G_0} \frac{d I}{d V}
= \cases{
1- \frac{3}{2}\left(\frac{eV}{T_K}\right)^2, 
& $eV\ll T_K$ \\
\frac{3\pi^2/16}{\ln^2(eV/T_K)}\,, 
& $eV \gg T_K$ 
}
\label{6.25}
\end{equation}
Thus, a large voltage bias has qualitatively the same destructive 
effect on the Kondo physics as the temperature does. The 
result~\eref{6.25} remains valid as long as $T\ll eV\ll \delta E$. 
If temperature exceeds the bias, $T\gg eV$, the differential 
conductance coincides with the linear conductance, see 
Equations~\eref{6.12} and \eref{6.21} above.

\section{Kondo effect in quantum dots with large spin}
\label{large_S}

If the dot's spin exceeds $1/2$~\cite{weis,Leo,Kogan}, then, as
discussed in Sec.~\ref{H_K} above, both exchange constants
$J_\gamma$ in the effective Hamiltonian~\eref{5.2} are finite and
positive. This turns out to have a dramatic effect on the dependence
of the conductance in the Kondo regime on temperature $T$
and on Zeeman energy $B$. Unlike the case of $S=1/2$ on the dot, 
see Fig.~\ref{overall}, now the dependences on $T$ and $B$ are
\textit{nonmonotonic}: initial increase of $G$ follows by a drop 
when the temperature is lowered~\cite{real,ISS} at $B=0$; the 
variation of $G$ with $B$ at $T=0$ is similarly nonmonotonic.
 
The origin of this peculiar behavior is easier to undersand by 
considering the $B$-dependence of the zero-temperature 
conductance~\cite{real}. We assume that the magnetic field 
$H_{\parallel}$ is applied \textit{in the plane} of the dot. Such
field leads to the Zeeman splitting $B$ of the spin states of the 
dot, see Eq.~\eref{6.17}, but barely affects the orbital motion of electrons.

At any finite $B$ the ground state of the system is not degenerate. 
Therefore, the linear conductance at $T = 0$ can be calculated from 
the Landauer formula
\begin{equation}
G = \frac{e^2}{h} \sum_s \left| \mathbb{S}_{s;RL}^2 \right|,
\label{7.1}
\end{equation}
which relates $G$ to the amplitude of scattering $\mathbb{S}_{s;RL}$ 
of an electron with spin $s$ from lead $L$ to lead $R$. The 
amplitudes $\mathbb{S}_{s;\alpha\alpha'}$ form a $2\times 2\,$ 
scattering matrix $\hat\mathbb{S}_s$. In the basis of  ``channels'', 
see Eq.~\eref{5.2}, this matrix is obviously diagonal, and its eigenvalues 
$\exp\negthinspace\left(2i\delta_{\gamma s}\right)$ are related to 
the scattering phase shifts $\delta_{\gamma s}$. The scattering 
matrix in the original $(R-L)$ basis is obtained from
\[
\hat\mathbb{S}_s = \hat U^\dagger 
{\rm diag} \left\{\rme^{2i\delta_{\gamma s}}\right\} 
\hat U,
\]
where $\hat U$ is a matrix of a rotation by an angle $\theta_0$, see Eq.~\eref{6.6}.
The Landauer formula~\eref{7.1} then yields
\begin{equation}
G = G_0 \frac{1}{2} \sum_s \sin^2\left(\delta_{1s} - \delta_{2s}\right),
\quad 
G_0 = \frac{2e^2}{h}\sin^2(2\theta_0) ,
\label{7.2}
\end{equation}
which generalizes the single-channel expression~\eref{6.14}. 

\Eref{7.2} can be further simplified for a particle-hole symmetric 
model~\eref{5.2}. Indeed, in this case the phase shifts satisfy 
$\delta_{\gamma \uparrow} + \delta_{\gamma\downarrow} = 0$,
cf.  Eq.~\eref{6.16}, which suggests a representation
\[
\delta_{\gamma s} = s\delta_\gamma .
\]
Substitution into~\eref{7.2} then results in
\begin{equation}
G = G_0 \sin^2\left(\delta_1 - \delta_2\right) .
\label{7.3}
\end{equation}

\begin{figure}[h]
\centerline{\includegraphics[width=0.7\textwidth]{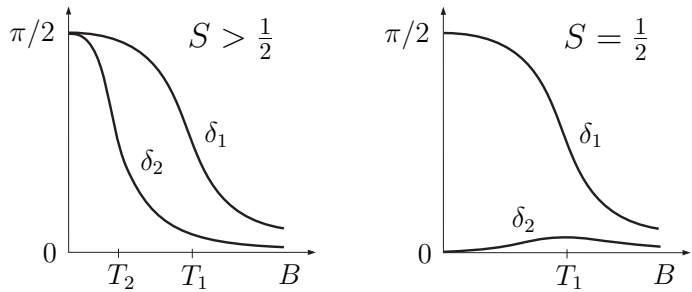}}
\vspace{2mm}
\caption{Dependence of the scattering phase shifts at the Fermi level 
on the magnetic field for $S>1/2$ (left panel) and $S=1/2$ (right panel).
}
\label{phase_shifts}
\end{figure}

\begin{figure}[h]
\centerline{\includegraphics[width=0.35\textwidth]{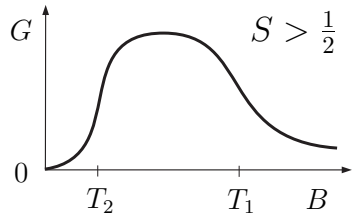}}
\vspace{2mm}
\caption{Sketch of the magnetic field dependence of the Kondo 
contribution to the linear conductance at zero temperature. The 
conductance as function of temperature exhibits a similar nonmonotonic 
dependence.}
\label{G_large_S}
\end{figure}

If spin on the dot $S$ exceeds $1/2$, then both channels of itinerant
electrons participate in the screening of the dot's spin~\cite{NB}.
Accordingly, in the limit $B\to 0$ both phase shifts $\delta_\gamma$ 
approach the unitary limit value $\pi/2$, see Fig.~\ref{phase_shifts}.  
However, the increase of the phase shifts on lowering the field is 
characterized by two different energy scales. These scales, the Kondo 
temperatures $T_1$ and $T_2$, are related to the corresponding 
exchange constants in the effective Hamiltonian~\eref{5.2},
\[
\ln\left(\frac{\Delta}{\,T_\gamma}\right) \sim \frac{1}{\nu J_\gamma} ,
\]
so that $T_1>T_2$ for $J_1 > J_2$. It is then obvious from Eq.~\eref{7.3} 
that the conductance across the dot is small both at weak $(B\ll T_2)$ 
and strong $(B\gg T_1)$ fields, but may become large $(\sim G_0)$ 
at intermediate fields $T_2\ll B \ll T_1$, see Fig.~\ref{G_large_S}. 
In other words, the dependence of zero-temperature conductance on 
the magnetic field is nonmonotonic.

This nonmonotonic dependence is in sharp contrast with the
monotonic increase of the conductance with lowering the field 
when $S = 1/2$. Indeed, in the latter case it is the channel whose 
coupling to the dot is the strongest that screens the dot's spin, 
while the remaining channel decouples at low energies~\cite{NB}, 
see Fig.~\ref{phase_shifts}. 

The dependence of the conductance on temperature $G(T)$ 
is very similar to $G(B)$\footnote{Note, however, that $ 
\langle \psi^\dagger_1(t)\psi_2^\pdag (t)
\psi_2^\dagger(0)\psi_1^\pdag(0)\rangle
\neq 
\langle \psi^\dagger_1(t)\psi_1^\pdag(0)\rangle
\langle \psi_2^\pdag(t) \psi_2^\dagger(0)\rangle
$ at finite $T$.
Therefore, unlike~\eref{6.14}, Eq.~\eref{6.8} does not allow 
for a simple generalization to the two-channel case.
}. 
For example, for $S = 1$ one obtains~\cite{real}
\begin{equation}
G/G_0 = \cases{
(\pi T)^2 \left(\frac{1}{T_1} - \frac{1}{T_2}\right)^2, & $T\ll T_2$ 
\\ 
\frac{\pi ^2}{2} 
\left[\frac{1}{\ln(T/T_1)} - \frac{1}{\ln(T/T_2)}\right]^2,  & $T\gg T_1$ 
}
\label{7.4}
\end{equation}
The conductance reaches its maximal value $G_{\rm max}$ 
at $T\sim\sqrt{T_1 T_2}$. The value of $G_{\rm max}$ can 
be found analytically for $T_1 \gg T_2$. For $S = 1$ 
the result reads~\cite{real}
\begin{equation}
G_{\rm max} = G_0 \left[1- \frac{3\pi^2}{\ln^2(T_1/T_2)}\right] .
\label{7.5}
\end{equation}

Consider now a Coulomb blockade valley with $N = {\rm even}$ 
electrons and spin $S=1$ on the dot. In a typical situation, the 
dot's spin in two neighboring valleys (with $N\pm 1$ electrons) is $1/2$. 
Under the conditions of applicability of the approximation~\eref{5.5}, 
there is a single non-zero exchange constant $J_{N\pm 1}$ for each 
of these valleys. If the Kondo temperatures $T_K$ are the same 
for both valleys with $S=1/2$, then $J_{N+1}=J_{N-1}=J_{\rm odd}$. 
Each of the two singly-occupied energy levels in the valley with $S=1$ 
is also singly-occupied in one of the two neighboring valleys. 
It then follows from Equations~\eref{5.4}-\eref{5.6} that the 
exchange constants $J_{1,2}$ for $S=1$ satisfy
\[
J_1 + J_2 = \frac{1}{2} \left(J_{N+1}+J_{N-1}\right) = J_{\rm odd}.
\]
Since both $J_1$ and $J_2$ are positive, this immediately implies that
$J_{1,2} < J_{\rm odd}$. 
Accordingly, both Kondo temperatures $T_{1,2}$ are expected 
to be smaller than $T_K$ in the nearby valleys with $S = 1/2$.

This consideration, however, is not applicable when the dot is 
tuned to the vicinity of the singlet-triplet transition in its ground 
state~\cite{Sasaki,induced_review,Leo,Kogan}, i.e. when the 
energy gap $\Delta$ between the triplet ground state and the singlet 
excited state of an isolated dot is small compared to the mean level 
spacing $\delta E$. In this case the exchange constants in the 
effective Hamiltonian~\eref{5.2} should account for additional 
renormalization that the system's parameters undergo when the 
high-energy cutoff (the bandwidth of the effective Hamiltonian) $D$ 
is reduced from $D\sim\delta E$ down to $D\sim \Delta\ll\delta E$~\cite{ST}, 
see also~\cite{PG}. The renormalization enhances the exchange 
constants $J_{1,2}$. If the ratio $\Delta/\delta E$ is sufficiently small, 
then the Kondo temperatures $T_{1,2}$ for $S=1$ 
may become of the same order~\cite{weis,Kogan}, or even 
significantly exceed~\cite{Sasaki,induced_review,Leo} the 
corresponding scale $T_K$ for $S=1/2$.

In GaAs-based lateral quantum dot systems the value of $\Delta$ can be
controlled by a magnetic field $H_\perp$ applied \textit{perpendicular
to the plane} of the dot~\cite{Leo}. Because of the smallness of the
effective mass $m^\ast$, even a weak field $H_\perp$ has a strong
orbital effect. At the same time, smallness of the quasiparticle g-factor 
in GaAs ensures that the corresponding Zeeman splitting remains 
small~\cite{induced_review}. Theory of the Kondo effect in
lateral quantum dots in the vicinity of the singlet-triplet transition
was developed in~\cite{ST_lateral}, see also~\cite{Zeeman}.

\section{Discussion}

In the simplest form of the Kondo effect considered in this review, 
a quantum dot behaves essentially as an artificial ``magnetic impurity'' 
with spin $S$, coupled via exchange interaction to two conducting 
leads. The details of the temperature dependence $G(T)$ of the linear 
conductance across a lateral quantum dot depend on the dot's spin~$S$. 
In the most common case of $S=1/2$ the conductance in the Kondo 
regime monotonically increases with the decrease of temperature, 
potentially up to the quantum limit $2e^2/h$. Qualitatively (although 
not quantitatively), this increase can be understood from the Anderson 
impurity model in which the dot is described by a single energy level. 
On the contrary, when spin on the dot exceeds $1/2$, the evolution of 
the conductance proceeds in two stages: the conductance first raises, 
and then drops again when the temperature is lowered.

In a typical experiment~\cite{kondo_exp}, one measures the 
dependence of the differential conductance on temperature $T$, 
Zeeman energy $B$, and dc voltage bias $V$. When one of these 
parameters is much larger than the other two, and is also large compared 
to the Kondo temperature $T_{\rm K}$, the differential conductance 
exhibits a logarithmic dependence
\begin{equation}
\frac{1}{G_0} \frac{dI}{dV}
\propto \left[\ln \frac{\mathrm{max} \left\{T,B,eV\right\}}{T_K} \right]^{-2},
\label{8.1}
\end{equation}
characteristic for the weak coupling regime of the Kondo system. 
Consider now a zero-temperature transport through a quantum dot 
with $S=1/2$ in the presence of a strong field $B\gg T_K$. In 
accordance with~\eref{8.1}, the differential conductance is small 
compared to $G_0$  both for $eV\ll B$ and for $eV\gg B$. However, 
the calculation in the third order of perturbation theory in the exchange 
constant yields a contribution that diverges logarithmically at 
$eV =B$~\cite{Appelbaum}. The divergence appears because at $eV=B$ 
the scattered electron has just the right amount of energy to allow for a 
real transition with a flip of spin. However, the full development of resonance 
is inhibited by a finite lifetime of the excited spin state of the dot~\cite{MWL,LW}. 
As a result, the peak in the differential conductance at $eV\sim B$ is broader 
and lower~\cite{MWL} then the corresponding peak at zero bias in the 
absence of the field. Even though for $eV\sim B\gg T_K$ the 
system is clearly in the weak coupling regime, a resummation of the 
perturbation series turns out to be a very difficult task, and the detailed 
shape of the peak is still unknown. This problem remains to be a subject 
of active research, see e.g.~\cite{RCPW} and references therein.

One encounters similar difficulties in studies of the effect of a weak 
ac excitation of frequency $\Omega\gtrsim T_K$ applied to the gate 
electrode~\cite{Elzerman} on transport across the dot. In close analogy 
with the usual photon-assisted tunneling~\cite{TG}, such perturbation is 
expected to result in the formation of satellites~\cite{HS} at 
$eV = n\hbar\Omega$ (here $n$ is an integer) to the zero-bias peak 
in the differential conductance. Again, the formation of the satellite peaks 
and the survival of the zero-bias peak in the presence of the ac excitation
are limited by the finite lifetime effects~\cite{KNG}.  

The spin degeneracy is not the only possible source of the Kondo 
effect in quantum dots. Consider, for example, a large dot connected 
by a single-mode junction to a conducting lead and tuned to the vicinity 
of the Coulomb blockade peak~\cite{KM}. If one neglects the finite 
level spacing in the dot, then the two almost degenerate charge state 
of the dot can be labeled by a pseudospin, while real spin plays the 
part of the channel index~\cite{KM,Matveev}. This setup turns out 
to be a robust realization~\cite{KM,Matveev} of the symmetric (i.e. 
having equal exchange constants) two-channel $S=1/2$ Kondo 
model~\cite{NB}. The model results in a peculiar temperature dependence 
of the observable quantities, which at low temperatures follow power 
laws with manifestly non-Fermi-liquid fractional powers.

It should be emphasized that in the usual geometry consisting of two 
leads attached to a small\footnote{i.e. with appreciable level spacing} 
Coulomb-blockaded quantum dot with $S=1/2$, 
only the conventional Fermi-liquid behavior can be observed at low 
temperatures. Indeed, in this case the two exchange constants in the 
effective exchange Hamiltonian~\eref{5.2} are vastly different, and 
their ratio is not tunable by conventional means, see the discussion in 
Sec.~\ref{H_K} above. A way around this difficulty was proposed 
recently in~\cite{OGG}. The key idea is to replace one of the leads in 
the standard configuration by a very large quantum dot, characterized 
by a level spacing $\delta E'$ and a charging energy $E_C^\prime$. 
At $T\gg \delta E'$, particle-hole excitations within this dot are allowed, 
and electrons in it participate in the screening of the smaller dot's spin. 
At the same time, as long as $T\ll E_C^\prime$, the number of electrons 
in the large dot is fixed. Therefore, the large dot provides for a separate 
screening channel which does not mix with that supplied by the remaining 
lead. In this system, the two exchange constants are controlled by the 
conductances of the dot-lead and dot-dot junctions. A strategy for tuning 
the device parameters to the critical point characterized by the two-channel 
Kondo physics is discussed in~\cite{2CK}. 

Finally, we should mention that the description based on the universal 
Hamiltonian \eref{2.12} is not applicable to large quantum dots subjected 
to a {\it quantizing} magnetic field $H_\perp$~\cite{checkers_1,checkers_2}. 
Such field changes drastically the way the screening occurs in a confined 
droplet of a two-dimensional electron gas~\cite{edge}. The droplet is 
divided into alternating domains containing compressible and incompressible 
electron liquids. In the metal-like compressible regions, the screening is almost 
perfect. On the contrary, the incompressible regions behave very much 
like insulators. In the case of lateral quantum dots, a large compressible 
domain may be formed near the center of the dot. This domain is surrounded 
by a narrow incompressible region separating it from another compressible 
ring-shaped domain formed along the edges of the dot~\cite{rings}. This 
system can be viewed as two concentric capacitively coupled quantum 
``dots" - the core dot and the edge dot~\cite{checkers_1,rings}. When 
the leads are attached to the edge dot, the measured conductance is sensitive 
to its spin state: if the number of electrons in the edge dot is odd, then
the conductance becomes large due to the Kondo effect~\cite{checkers_1}. 
Changing the field causes redistribution of electrons between the core and 
the edge, resulting in a striking checkerboard-like pattern of high- and 
low-conductance regions~\cite{checkers_1,checkers_2}. This behavior 
persists as long as the Zeeman energy remains small compared to the 
Kondo temperature. Note that compressible regions are also formed 
around an {\it antidot} -- a potential hill in a two-dimensional electron 
gas in the quantum Hall regime~\cite{GS}. Both Coulomb blockade 
oscillations and Kondo-like behavior were observed in these 
systems too~\cite{antidot}.

\section{Summary}

Kondo effect arises whenever a coupling to a Fermi gas induces
transitions within otherwise degenerate ground state multiplet of an
interacting system. Both coupling to a Fermi gas and interactions are
naturally present in a nanoscale transport experiment. At the same
time, many nanostructures can be easily tuned to the vicinity of a
degeneracy point. This is why the Kondo effect in its various forms 
often influences the low temperature transport in meso- and nanoscale
systems. 

In this article we reviewed the theory of the Kondo effect in transport 
through quantum dots. A Coulomb-blockaded quantum dot behaves 
in many aspects as an artificial ``magnetic impurity'' coupled via exchange 
interaction to two conducting leads. Kondo effect in transport through 
such ``impurity'' manifests itself in the lifting of the Coulomb blockade 
at low temperatures, and, therefore, can be unambiguously identified. 
Quantum dot systems not only offer a direct access to transport properties 
of an artificial impurity, but also provide one with a broad arsenal of tools 
to tweak the impurity properties, unmatched in conventional systems. The 
characteristic energy scale for the intra-dot excitations is much smaller 
than the corresponding scale for natural magnetic impurities. This allows 
one to induce degeneracies in the ground state of a dot which are more 
exotic than just the spin degeneracy. This is only one out of many possible 
extensions of the simple model discussed in this review.

\ack
The research at the University of Minnesota was supported by 
NSF grants DMR02-37296, and EIA02-10736.



\Bibliography{99}

\bibitem{blockade}
Kouwenhoven L P \etal 
1997 {\it Mesoscopic Electron Transport} 
ed Sohn L L \etal
(Dordrecht: Kluwer) p~105
\nonum
Kastner M A 
1992 \RMP {\bf 64} 849
\nonum
Meirav U and Foxman E B 
1996 {\it Semicond. Sci. Technol.} {\bf 11} 255\nonum
Kouwenhoven L P and Marcus C M
1998 {\it Phys. World} {\bf 11} 35

\bibitem{Devoret}
Joyez P \etal 1997 \PRL {\bf 79} 1349
\nonum
Devoret M and Glattli C 
1998 {\it Phys. World} {\bf 11} 29

\bibitem{kondo_exp}  
Goldhaber-Gordon D {\etal}
1998 {\it Nature} {\bf 391} 156
\nonum
Cronenwett S M, Oosterkamp T H and Kouwenhoven L P 
1998 {\it Science} {\bf 281} 540 
\nonum
Schmid J \etal
1998 {\it Physica B} {\bf 256-258} 182

\bibitem{kondo_popular}
Kouwenhoven L and Glazman L 
2001 {\it Physics World} {\bf 14} 33

\bibitem{Kondo} 
Kondo J 
1964 {\it Prog. Theor. Phys.} {\bf 32} 37

\bibitem{vertical} 
Tarucha S \etal 
2000 \PRL {\bf 84} 2485
\nonum
Kouwenhoven L P, Austing D G and Tarucha S 
2001 {\it Rep. Prog. Phys.} {\bf 64} 701

\bibitem{Sasaki}
Sasaki S \etal 
2000 {\it Nature} {\bf 405} 764

\bibitem{induced_review}
Pustilnik M \etal 
2001 {\it Lecture Notes in Physics} {\bf 579} 3 
(Pustilnik M \etal {\it Preprint} cond-mat/0010336)

\bibitem{nanotube}
Nyg{\aa}rd J, Cobden D H and Lindelof P E 
2000 {\it Nature} {\bf 408} 342
\nonum
Liang W, Bockrath M and Park H 
2002 \PRL {\bf 88} 126801 

\bibitem{Park} 
Park J \etal 
2002 {\it Nature} {\bf 417} 722
\nonum
Liang W \etal 
2002 {\it Nature} {\bf 417} 725

\bibitem{mirage}
Knorr N \etal 
2002 \PRL {\bf 88} 096804
\nonum
Manoharan H C \etal 
2000 {\it Nature} {\bf 403} 512
\nonum
Madhavan V \etal 
1998 {\it Science} {\bf 280} 567
\nonum
Chen W \etal 
1999 \PRB {\bf 60} R8529
\nonum
Li J T \etal 
1998 \PRL {\bf 80} 2893

\bibitem{RMT1} 
Berry M V 1985 {\it Proc. R. Soc. A} {\bf 400} 229
\nonum
Altshuler B L and Shklovskii B I 1986 {\it Sov. Phys. JETP} {\bf 64} 127

\bibitem{RMT_reviews}
Beenakker C W J 1997 \RMP {\bf 69} 731
\nonum
Alhassid Y 2000 \RMP {\bf 72} 895

\bibitem{Mehta}
Mehta M L 1991 {\it Random Matrices} (New York: Academic Press)

\bibitem{RMT} 
Altshuler B L \etal
1997 \PRL {\bf 78} 2803
\nonum 
Agam O \etal 
1997 \PRL {\bf 78} 1956
\nonum
Blanter Ya M 1996 \PRB {\bf 54} 12807
\nonum
Blanter Ya M and Mirlin A D 
1998 \PRB {\bf 57} 4566 
\nonum
Blanter Ya M, Mirlin A D and Muzykantskii B A
1997 \PRL {\bf 78} 2449
\nonum
Aleiner I L and Glazman L I 1998 \PRB {\bf 57} 9608

\bibitem{KAA}
Kurland I L, Aleiner I L and Altshuler B L 
2000 \PRB {\bf 62} 14886

\bibitem{ABG}
Aleiner I L, Brouwer P W and Glazman L I 
2002 {\it Phys. Rep.} {\bf 358} 309

\bibitem{wave_functions}
Berry M V 1977 \JPA {\bf 10} 2083
\nonum
Blanter Ya M and Mirlin A D 1997 \PR {\it E} {\bf 55} 6514
\nonum
Blanter Ya M, Mirlin A D and Muzykantskii B A
2001 \PRB {\bf 63} 235315
\nonum
Mirlin A D 2000 {\it Phys. Rep.} {\bf 326} 259

\bibitem{Ziman}
Ziman J M 
1972 {\it Principles of the Theory of Solids}
(Cambridge: Cambridge University Press) 

\bibitem{spin}  
Brouwer P W, Oreg Y and Halperin B I
1999 \PRB {\bf 60} R13977
\nonum
Baranger H U, Ullmo D and Glazman L I 
2000 \PRB {\bf 61} R2425

\bibitem{spin_exp}
Potok R M \etal 
2003 \PRL {\bf 91} 016802 
\nonum
Folk J A 
2001 {\it Phys. Scripta} {\bf T90} 26 
\nonum
Lindemann S \etal
2002 \PRB {\bf 66} 195314

\bibitem{KM} 
Matveev K A 1995 \PRB {\bf 51} 1743

\bibitem{KF} 
Flensberg K 1993 \PRB {\bf 48} 11156

\bibitem{real}
Pustilnik M and Glazman L I 2001 \PRL {\bf 87} 216601

\bibitem{rate}
Kulik I O and Shekhter R I  1975 {\it Sov. Phys.--JETP} {\bf 41} 308
\nonum 
Glazman L I and Shekhter R I 1989 \JPCM {\bf 1} 5811

\bibitem{Giaever}
Giaever I and Zeller H R 1968 \PRL {\bf 20} 1504
\nonum 
Zeller H R and Giaever I 1969 \PR {\bf 181} 789

\bibitem{AN}
Averin D V and Nazarov Yu V 1990 \PRL {\bf 65} 2446

\bibitem{Abrikosov} 
Abrikosov A A 1988 {\it Fundamentals of the Theory of Metals}
(Amsterdam: North-Holland) 

\bibitem{AG}
Aleiner I L and Glazman L I 1996 \PRL {\bf 77} 2057

\bibitem{Gefen}
Silva A, Oreg Y, and Gefen Y 2002
\PRB {\bf 66} 195316
\nonum
Baltin R and Gefen Y 1999 \PRL {\bf 83} 5094

\bibitem{AM}
Glazman L I and Raikh M E 1988 {\it JETP Lett.} {\bf 47} 452 (1988)
\nonum 
Ng T K and Lee P A 1988 \PRL {\bf 61} 1768

\bibitem{unitary}
van der Wiel W G \etal 
2000 {\it Science} {\bf 289} 2105
\nonum
Ji Y, Heiblum M and Shtrikman H 
2002 \PRL {\bf 88} 076601

\bibitem{Erice}
Glazman L I and Pustilnik M 
2003 {\it New Directions in Mesoscopic Physics (Towards Nanoscience)} 
ed Fazio R \etal (Dordrecht: Kluwer) p~93
(Glazman L I and Pustilnik M {\it Preprint} cond-mat/0302159)

\bibitem{old_reviews} 
Duke C B 1969{\it Tunneling in Solids} (New York: Academic Press)
\nonum
Rowell J M 1969 {\it Tunneling Phenomena in Solids} 
ed Burstein E and Lundqvist S (New York: Plenum Press) p~385

\bibitem{classics} 
Wyatt A F G 1964 \PRL {\bf 13} 401
\nonum
Logan R A and Rowell J M 1964 \PRL {\bf 13} 404

\bibitem{Appelbaum} 
Appelbaum J 1966 \PRL {\bf 17} 91
\nonum
Appelbaum J A 1967 \PR {\bf 154} 633

\bibitem{Anderson}
Anderson P W 1966 \PRL {\bf 17} 95

\bibitem{Fiete} 
Fiete G A \etal 
2002 \PRB {\bf 66} 024431

\bibitem{SW}
Schrieffer J R and Wolff P A 
1966 \PR {\bf 149} 491

\bibitem{SI}
Silvestrov P G and Imry Y 2000 \PRL {\bf 85} 2565

\bibitem{MWL}
Meir Y, Wingreen N S and Lee P A 1993 \PRL {\bf 70} 2601

\bibitem{NB} 
Nozi\`{e}res P and Blandin A 1980 {\it J. Physique} {\bf 41} 193

\bibitem{Kondo_reviews}
Coleman P 
2002 {\it Lectures on the Physics of Highly Correlated Electron Systems} 
ed Mancini F (New York: American Institute of Physics) p~79
(Coleman P {\it Preprint} cond-mat/0206003)
\nonum
Hewson A S 
1997 {\it The Kondo Problem to Heavy Fermions} 
(Cambridge: Cambridge University Press)

\bibitem{PWA_book} 
Anderson P W 1997 {\it Basic Notions of Condensed Matter Physics} 
(Reading: Addison-Wesley)
\bibitem{Wilson}
Wilson K G 1975 \RMP{\bf 47} 773

\bibitem{GHL} 
Glazman L I, Hekking F W J and Larkin A I 
1999 \PRL {\bf 83} 1830 

\bibitem{bethe} 
Tsvelick A M and Wiegmann P B 1983 {\it Adv. Phys.} {\bf 32} 453
\nonum
Andrei N, Furuya K and Lowenstein J H 1983 \RMP {\bf 55} 331 

\bibitem{AL} 
Affleck I and Ludwig A W W 1993 \PRB {\bf 48} 7297

\bibitem{PG} 
Pustilnik M and Glazman L I 2001 \PRB {\bf 64} 045328

\bibitem{AAA}  
Abrikosov A A 1965 {\it Physics} {\bf 2} 5
\nonum
Abrikosov A A 1969 {\it Sov. Phys.--Uspekhi} {\bf 12} 168

\bibitem{Newton} 
Newton R G 2002 {\it Scattering Theory of Waves and Particles} 
(Mineola: Dover)

\bibitem{PWA}  
Anderson P W 1970 \JPC {\bf 3} 2436

\bibitem{AYH}
Anderson P W, Yuval G and Hamann D R 
1970 \PRB {\bf 1} 4464 

\bibitem{N} 
Nozi\`{e}res P 1974 {\it J. Low Temp. Phys.} {\bf 17} 31
\nonum 
Nozi\`{e}res P 1978 {\it J. Physique} {\bf 39} 1117

\bibitem{Costi}
Costi T A, Hewson A C and Zlati\'{c} V 1994 \JPCM {\bf 6} 2519

\bibitem{weis} 
Schmid J \etal 
2000 \PRL {\bf 84} 5824 

\bibitem{Leo}
van der Wiel W G 
2002 \PRL {\bf 88} 126803

\bibitem{Kogan} 
Kogan A \etal 
2003 \PRB {\bf 67} 113309

\bibitem{ISS}
Izumida W, Sakai O and Shimizu Y
1998 {\it J. Phys. Soc. Jpn.} {\bf 67} 2444 

\bibitem{ST} 
Eto M and Nazarov Yu V 2000 \PRL {\bf 85} 1306
\nonum 
Pustilnik M and Glazman L I  2000 \PRL {\bf 85} 2993

\bibitem{ST_lateral}
Golovach V N and Loss D 2003 {\it Europhys. Lett.} {\bf 62} 83
\nonum
Pustilnik M, Glazman L I and Hofstetter W
2003 \PRB {\bf 68} 161303(R)   

\bibitem{Zeeman} 
Pustilnik M, Avishai Y, and Kikoin K 2000 \PRL {\bf 84} 1756

\bibitem{LW} 
Losee D L and Wolf E L 
1969 \PRL {\bf 23} 1457

\bibitem{RCPW}
Rosch A, Costi T A, Paaske J and W{\"o}lfle P
2003 \PRB {\bf 68} 014430
\nonum
Paaske J, Rosch A and W{\"o}lfle P
{\it Preprint} cond-mat/0307365

\bibitem{Elzerman}
Elzerman J M \etal 
2000 {\it J. Low Temp. Phys.} {\bf 118} 375

\bibitem{TG}
Tien P K and Gordon J P 
1963 \PR 129, 647 (1963)

\bibitem{HS}
Hettler M H and Schoeller H 1995 \PRL {\bf 74} 4907 

\bibitem{KNG} 
Kaminski A, Nazarov Yu V and Glazman L I 
1999 \PRL {\bf 83} 384

\bibitem{Matveev}
Matveev K A  1991 {\it Sov. Phys. JETP} {\bf 72} 892

\bibitem{OGG}
Oreg Y and Goldhaber-Gordon D 2003
\PRL {\bf 90} 136602 

\bibitem{2CK}
Pustilnik M \etal 
2004 \PRB {\bf 69} 115316

\bibitem{checkers_1}
Keller M \etal 
2001 \PRB {\bf 64} 033302 
\nonum
Stopa M \etal 
2003 \PRL {\bf 91} 046601 

\bibitem{checkers_2}
Maurer S M \etal 
1999 \PRL {\bf 83} 1403 
\nonum
Sprinzak D \etal 
2002 \PRL {\bf 88} 176805 
\nonum 
F{\"u}hner C \etal 
2002 \PRB {\bf 66} 161305(R) 
\nonum
Keyser U F \etal 
2003 \PRL {\bf 90} 196601

\bibitem{edge}
Beenakker C W J 
1990 \PRL {\bf 64} 216
\nonum 
Chang A M 
1990 {\it Solid State Commun.} {\bf 74} 871
\nonum
Chklovskii D B, Shklovskii B I and Glazman L. I. 
1992 \PRB {\bf 46} 4026

\bibitem{rings}
McEuen P L \etal
1992 \PRB {\bf 45} 11 419
\nonum
Evans A K, Glazman L I and Shklovskii B I
1993 \PRB {\bf 48} 11 120

\bibitem{GS}
Goldman V J and Su B 1995 {\it Science} {\bf 267} 1010

\bibitem{antidot}
Kataoka M \etal 1999 \PRL {\bf 83} 160
\nonum
Kataoka M \etal 
2002 \PRL {\bf 89} 226803

\endbib
\end{document}